\documentclass[11pt,a4paper]{article}
\pdfoutput=1
\usepackage[utf8]{inputenc}
\usepackage[english]{babel}
\usepackage{extarrows}
\usepackage{amsmath}
\usepackage{amsfonts}
\usepackage{amssymb}
\usepackage{graphicx}
\usepackage{fourier}

\usepackage{caption}
\usepackage{subcaption}
\usepackage{float}
\usepackage{color}
\usepackage{abstract}
\usepackage{appendix}
\usepackage{authblk}
\usepackage{xcolor}

\numberwithin{equation}{section}

\usepackage[hidelinks]{hyperref}

\newcommand\be{\begin{equation}}
\newcommand\ee{\end{equation}}
\newcommand\ba{\begin{eqnarray}}    
\newcommand\ea{\end{eqnarray}}      

\title{Pacman geometries and the Hayward term in JT gravity}

\author[a]{Ra\'ul Arias}
\author[a]{Marcelo Botta-Cantcheff}
\author[b]{Pedro J. Martinez}

\affil[a]{Instituto de F\'isica La Plata - CONICET and 

Departamento de F\'isica, Universidad Nacional de La Plata 

C.C. 67, 1900, La Plata, Argentina}

\affil[b]{Centro At\'omico Bariloche, San Carlos de Bariloche, Argentina}

\date{}
\usepackage[left=2cm,right=2cm,top=2cm,bottom=2cm]{geometry}

\begin{document}
  
\maketitle
\thispagestyle{empty}

\begin{abstract}

We study the Hayward term describing corners in the boundary of the geometry in the context of the Jackiw-Teitelboim  gravity. These corners naturally arise in the computation of Hartle-Hawking wave functionals and reduced density matrices, and give origin to AdS spacetimes with conical defects.

This set up constitutes a lab to manifestly realize many aspects of the construction recently proposed in \cite{Botta2020}. In particular, it can be shown that the Hayward term is required to reproduce the flat spectrum of R\'enyi entropies in the Fursaev's derivation, and furthermore, the action with an extra Nambu-Goto term associated to the Dong's cosmic brane prescription appears naturally.

On the other hand, the conical defect coming from Hayward term contribution are subtly different from the defects set as pointlike \emph{sources} studied previously in the literature. We study and analyze these quantitative differences in the path integral and compare the results.
Also study previous proposals on the superselection sectors, and by computing the density operator we obtain the Shannon entropy and some novel results on the symmetry group representations and edge modes. It also makes contact with the so-called \emph{defect operator} found in \cite{Jafferis2019}.

Lastly, we obtain the area operator as part of the gravitational modular Hamiltonian, in agreement with  the Jafferis-Lewkowycz-Maldacena-Suh proposal.

\end{abstract}

\newpage

\tableofcontents

\section{Introduction}

In recent years Jackiw-Teitelboim gravity \cite{Jackiw1984, Teitelboim1983} has been a fruitful field of study to probe many ideas related to information loss in Black Holes and realizing models where explicit calculations are manageable. In particular it has been capable of describe the conjectured Page curve for the BH entropy \cite{Penington2019, Marolf2019, Mahajan2019, Prem2020}. Many studies and models on entanglement entropy as well as properties of the partitions functions have been exhaustively studied in this context \cite{Jafferis2019, Lin2018, Lin2021, Saad2019}, and the question of the holographic correspondence with a boundary theory has also received a lot of attention \cite{Kitaev2015, Polchinski2016, Maldacena2016}. Some original proposals can be put in practice in JT that hopefully can be generalized to higher dimensions. 

\begin{figure}
\begin{subfigure}{0.49\textwidth}
\centering
\includegraphics[width=.9\linewidth] {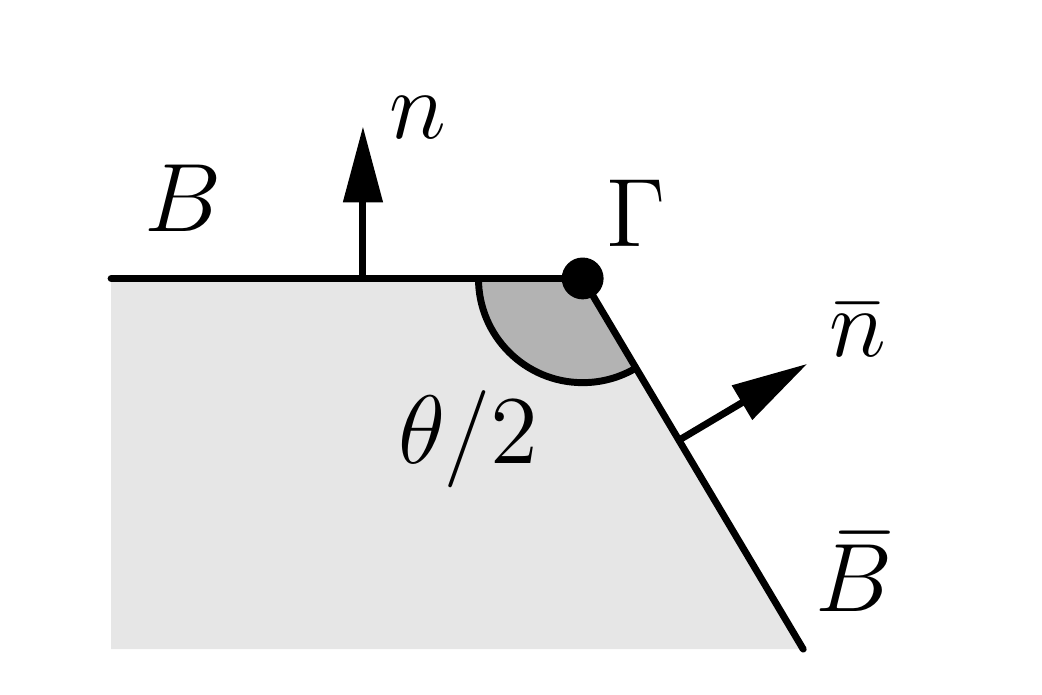}
\end{subfigure}
\begin{subfigure}{0.49\textwidth}
\centering
\includegraphics[width=.9\linewidth] {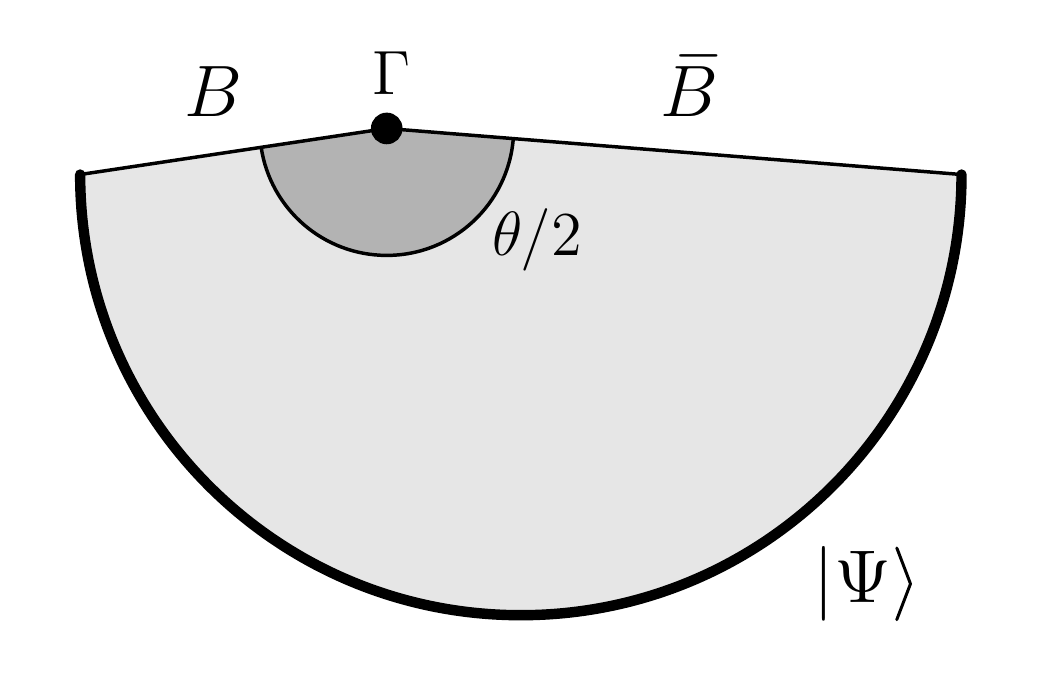}
\end{subfigure}
  \caption{(a) A standard situation where a Hayward boundary term must be considered. The spacetime has a boundary with a non-smooth corner at $\Gamma=B\cap \bar B$. The Hayward term can be expressed in terms of the internal angle $\pi-\theta/2=\cos^{-1}\, (n \cdot  \bar n)$. (b) We consider an Euclidean path integral Hartle-Hawking state $|\Psi\rangle$ projected into a non smooth basis characterized by a region $\Sigma=B\bigcup\bar B$. } 
  \label{fig:H2}
\end{figure}

Spacetimes whose boundary is not smooth and present corners \emph{require} to add a Hayward term in addition to the typical Gibbons-Hawking-York boundary term in order to have a well posed variational problem in gravity \cite{Hayward1993}. If there is a codim-2 corner $\Gamma$, see Fig. \ref{fig:H2}(a), that splits the boundary $\Sigma$ in two smooth components 
 $B$ and $\bar B$ with respective normal vectors $n$ and $\bar n$, thus the standard gravitational action has an extra term given by
$$
I_H^{d>2}\equiv\frac{1}{8\pi G}\int_\Gamma  \sqrt{\gamma}\; \cos^{-1}\, (n \cdot \bar n),   
$$
where $\gamma$ is the induced metric on $\Gamma$. A study of the effects of such a term for $d>2$ was carried out in \cite{Botta2020,Takayanagi2019}. Two dimensional JT gravity is a very convenient lab to test the consequences of this idea, the area element shall be substituted by $\Phi_\Gamma$, the Dilaton field in a point $\Gamma$ following an implicit standard dimensional reduction scheme. Strictly, in JT gravity the Hayward term takes the form
\be
I_H^{d=2} \equiv \frac{1}{8\pi G}\; \cos^{-1}\, (n \cdot  \bar n) \;\Phi_\Gamma.\nonumber
\ee  
Even though the Hartle-Hawking wave functional and (reduced) density matrix elements are described by dominant geometries whose boundaries generically involve corners,
surprisingly enough, the Hayward term has still been very little studied in the JT literature.
 Precisely one of the main goals of this manuscript is to probe, in the specific example of JT gravity, the holographic prescriptions using this term to compute R\'enyi entropies and modular flow \cite{Botta2020,Takayanagi2019}, in contrast with standard cosmic brane proposals \cite{Dong2016, Dong2018, Almheiri2019, Ellerin2021}.

In the context of the holographic duality, the entanglement entropy of ordinary QFT in a subregion of the boundary is given by a quarter of the area of a minimal surface embedded in the bulk spacetime \cite{Ryu2006}, capturing the Bekenstein-Hawking law for the black holes entropy \cite{Bekenstein1973, Hawking1975} as a particular case. This rule can also be generalized to a suitable one-parameter extension of the von-Neumann entropy $\hat{S}_n$ called refined R\'enyi entropies, which is a quarter of the area of a \emph{cosmic brane} minimally coupled with gravity with a tension \cite{Dong2016}
\be\label{tension-dongC}
T_n = \frac{n-1}{4n\,G}\;\;.
\ee
However, viewing this prescription as an entirely holographic model, the cosmic brane plays no natural role in the holographic duality. Thus, a derivation within a theory of pure gravity becomes of interest to the holographic dictionary. In the original prescription \cite{Dong2016} this can be achieved coming from a saddle of a replicated boundary condition, and the cosmic brane appears as an auxiliary object that effectively sources the correct dual geometry. Moreover, this construction lacks a description at the level of states and the density matrix.

This issue was solved in \cite{Botta2020} by observing that the Hartle-Hawking wave-functionals generically contains codim-2 corner contributions in the form of Hayward terms for non-smooth choices of the initial space slice, see Fig. \ref{fig:H2}(b). The geometry can be completed to compute the saddle contributions 
to the partition function where a conical geometry appears with the correct deficit angle predicted by the Dong's recipe. 
It has also been shown that such term can explain the Ryu-Takayanagi law \cite{Takayanagi2019}.

One of the advantages of the present approach is that the density matrix captures this information in the corner term (proportional to the codim-2 area), thus explaining that the Jafferis-Lewkowycz-Maldacena-Suh (JLMS) \cite{JLMS} proposal for the gravitational modular Hamiltonian explicitly contains the area operator \cite{Botta2020}. As a result, we will compute the density operator and show this claim in the JT example.
It is worth emphasizing here that the cosmic brane model \cite{Dong2016} is equivalent to the  proposal 
based on the Hayward/corner term at the level of the partition function, whose saddles are  closed, periodic, Euclidean geometries, but the last model furthermore provides the underlying description of the reduced density matrix and modular Hamiltonian in gravity. Moreover this is better supported by holographic arguments on the states and their matrix elements.

The gravitational representation of the states are the Hartle-Hawking wave-functionals, described as Euclidean path integrals with specific conditions on the asymptotic boundaries, complemented with initial conditions on an \emph{arbitrary} spacelike  (initial) surface $\Sigma$ that in general, may include a corner described by a non-trivial Hayward term in the action whose saddle geometry is depicted in Fig. \ref{fig:H2}(b). Then, by gluing two of these geometries on one of the two sides corresponding to the complement of the entanglement wedge, we obtain a Pacman geometry representing the gravitational (reduced) density matrix of a holographic state, see Fig. \ref{fig:Pacman}(a). 

We are going to focus our approach in these Pacman geometries, $M_P$, 
which are Euclidean spacetimes with specific (possibly arbitrary) boundary data on its internal boundaries $B^\pm$ and whose action necessarily contains a Hayward term describing the corner contribution in order to have a well posed problem under Dirichlet boundary data \cite{Hayward1993,Takayanagi2019}.
An analogous study is highly non-trivial in higher dimensions due to the many dofs living in the $B^\pm$ surfaces. We will find that the JT examples allow to properly define these dofs. 
 
Moreover, Pacman geometries are precisely the geometries required to compute the matrix elements of the reduced gravitational density matrix trough a path integral
\be\label{rhoB-MP}  \langle B^+ |\rho_r |B^-\rangle \; \sim  \; e^{-I[M_P]} + \ldots \approx \, e^{-I_{bulk} [M_P]} \, e^{-I_{GHY}[\partial M_P , B^\pm ] + \frac{(2\pi - \theta) }{8\pi G}a(\Gamma_{})} + \ldots\; , 
\ee
where $\dots$ represent subleading contributions to the path integral. The $\partial M_P$ denotes the asymptotic boundary and the last term is the Hayward term due to the corner with opening angle $\theta$. In JT gravity $a$ is given by the value of the Dilaton field $\Phi$ at the point $\Gamma$. The subindex in $\rho_r$ is schematic for the moment: it labels the irrep (superselection sector) of the density matrix, that we will claim to be associated to $\Gamma$ and  $a(\Gamma)$ and which will be properly introduced in the main body of the paper.

 Finally, by taking the trace of this density matrix we can obtain an entirely gravitational partition function with a conical defect described by the remaining Hayward term contribution. Since the origin of this contribution lies in the \emph{boundary} rather in the bulk interior (as in approaches \cite{Mertens2019, Witten2020a, Witten2020b, Mefford2020}), there are subtly but important differences in the nature and description of the defect and in the computations. For instance, there are many radical implications on the computation of the R\'enyi entropies and derivations of the Dong's construction.  

There is another paradigm that we are going to test in the present manuscript
related to the structure of \emph{sectors}, or blocks, such that the density matrix can be represented in the gravity side. In a supplementary proposal done in \cite{Botta2020}, subtly different to the notion of \emph{fixed area states} proposed in \cite{Dong2018}, the idea is that each codim-2 surface $\Gamma$, whose area is fixed, splits the space in two regions such that $\Gamma$ works as the entangling surface for gravitational dofs. This will be discussed more deeply in Sec. \ref{bipartite} and the consequences of such statement will be tested through the paper. This will shed light on the splitting of dofs under holographic mapping, the so called \emph{factorization map} and the edge modes.

There are other aspects related to this discussion that arise when gravity is treated as a gauge theory and in particular on the presence of edge modes because of the entangling surface $\Gamma$ bounding the spacetime dofs \cite{Lin2018,Jafferis2019}. Edge modes can be understood as dofs emerging from pure gauge ones as a bounding surface is imposed. In \cite{Takayanagi2019}, the authors also pointed out and studied the connection between the Hayward term and edge modes in gravity, and we will go some steps further in this issue.

The paper is organized as follows. In Sec. \ref{bipartite} we show how the holographic factorization proposed in \cite{Botta2020} works for this two dimensional example leading to the notion of fixed area sectors and an interpretation of the RT formula that is directly related to edge modes. In Sec. \ref{HayenJT} we will show explicitly how the Hayward term appears in JT gravity by varying the action and looking for the boundary terms that contribute to it. We will also obtain the on-shell solution to the system. In Sec. \ref{sources} we study the partition function obtained by taking the trace of the density matrix for the Pacman geometry. We make explicit the similarities and differences with the previous results about JT gravity with defects. We also obtained the spectral density. In Sec. \ref{replica} and \ref{modular} we compute the R\'enyi and refined R\'enyi entropy from the partition functions obtained before and give a derivation of the JLMS in this context respectively. In Sec. \ref{livingontheedge} we add a brief commentary on the edge modes by writing an action that takes them in account and constructing the conjugate pairs and their commutation relation. Finally we write a summary of results and conclusions in Sec. \ref{conclu}.

\subsection*{List of achievements and results}

\begin{itemize}
    \item Ideas and computations of the proposal of gravity with Hayward term of \cite{Botta2020} were tested for JT gravity.

\item Supplementary suppositions of \cite{Botta2020} in the JT lab on the holographic correspondence between subsystems were investigated. In particular, the von Neumann decomposition of the bulk Hilbert space in superselection (SS) sectors, claimed to be labeled by the entangling surfaces (a \emph{point} in 2d) $\Gamma$. These sectors are characterized by observables as the areas $\Phi_\Gamma$ and we compare this point of view with the fixed area states \cite{Dong2018}.
   
\item In this sense, a relation between these SS sectors and \emph{representations} as in ordinary gauge theories are found as well as several remarks on the symmetry group and its representations; and in particular on \emph{edge modes}. We recover the result of \cite{Lin2018} using different arguments and propose a  
generalization of her formula in presence of a conical defect, which indirectly implies that the symmetry/representations are deformed in this case. 
     
\item The partition function for this theory of gravity with corner terms is evaluated, at classical and quantum level.
    
\item The density of states of an hypothetical dual random matrix model, $\Omega(E)$ is computed.
    
\item A novel formula for the Euclidean JT action is found as the conical geometry is recovered by closing the Pacman manifold $M_P$. We argue that this is crucial to compute R\'enyi entropies correctly.
    
\item Well-established recipes \cite{Fursaev2006, Dong2016} to compute R\'enyi entropies are reproduced.

\item The \emph{defect operator} introduced in \cite{Jafferis2019} is also reproduced.

\item The modular Hamiltonian in JT gravity (+ corner terms) is computed and the JLMS proposal is reproduced.

\end{itemize}

\section{Bi-partite QM systems and holographic factorization}\label{bipartite}


The area laws and the extremality of RT/HRT surfaces must arise from a suitable semiclassical approximation of gravity as leading contributions; thus, the definition of spacetime regions as the entanglement wedge associated to a subset of the boundary dofs 
is unclear at quantum level. Moreover, the precise holographic correspondence between dof subsets is yet unknown.
The $AdS_{2}/CFT$ version of holography can be auspicious to study this.

Since aAdS$_{1+1}$ has two asymptotic boundaries $L,R$, one can compute entanglement quantities between both sides. They should be dual to a quantum mechanical system. 

Based on the same arguments that \cite{Botta2020}, our prescription here is that the reduced density matrix
of the quantum system $L$ (or $R$), has a block diagonal structure
\be \label{rhoA-osum} \rho(L) = \bigoplus_{\Gamma}\; \rho (\Gamma) \;. \ee
The physical interpretation of this expression is that given a fixed splitting of the QFT degrees of freedom living on the boundary in two subsets $A \cup \bar A$, then one should consider a \emph{sum} over all the possible splittings on the bulk dofs separated by a codim-2 (entangling) surface $\Gamma$ such that $\partial \Gamma =\partial A$.
A remarkable ingredient of this proposal is that one must view each bulk-splitting as a different superselection (SS) sector or \emph{representation}, and consider the direct sum over them.

This resembles the von Neumann's theorem, see e.g. Appendix of \cite{HarlowReview}.
Since the algebra of operators in the quantum theory defined on the boundary can be assumed to be a von Neumann algebra \cite{Lashkari2018}, then in the context of the gauge/gravity duality, both Hilbert spaces can be equated upon a von Neumann decomposition of the bulk Hilbert space as \cite{Harlow2018, Botta2020}
 \be\label{dsumH} 
 {\cal H}_L\otimes_{} {\cal H}_{R} \equiv \,\bigoplus_\Gamma \; {\cal H}_B\otimes {\cal H}_{\bar B}\,.
 \ee
So the reduced density matrix \eqref{rhoA-osum} results from taking the partial trace on ${\cal H}_R$ of the global state, while in the bulk one must sum over all possible entangling surfaces. 

Such a structure becomes more evident by considering a formal discretization of $\Sigma$ in a finite set of (in 1+1) points $\Sigma_M = \{q_1 \dots q_M\}$, so that the HH wave function has $M$ variables, and the state belongs to a finite dimensional Hilbert space dim$ {\cal H}_M = M$. In principle, one could also add independent fields to live on this lattice, but we disregard this possibility here to provide a minimal example. 

Consider a splitting taking place at any of these points,
say $\,\Gamma \equiv q_k$, 
the density matrix of the gravitational subsystem $B = \{q_1\dots q_k\} \subset \Sigma_M$, is a $k\times k$\emph{ matrix}
\be 
\rho(\Gamma) = p(k) \, \rho_{ab}(k)  \qquad\qquad a,b=1, \dots k \nonumber.
\ee
This is nothing but a $k$-\emph{dimensional representation} of the state $\rho(L)$, 
where the prefactor $p(k) \sim \, e^{-c\Phi(q_k) / G}$ is a number interpreted as the probability of the representation that shall be obtained within the gravity theory. Therefore, a complete description of the state of this system shall be the direct sum of \emph{blocks}
\be \label{rhoA-osum-discrete} \rho (L) = \bigoplus^{M-1}_{k=1}\; p_k\,\rho (k), \ee  which is the discrete version of \eqref{rhoA-osum}.

The so-called factorization problem in JT gravity is closely related to this decomposition and has already been discussed in detail \cite{Harlow2018}, but this is not our goal here and it remains for future research. For the most of applications studied in this work, we are interested in the formula to compute the partition function (and the entropy) in the field theory in terms of the theory of gravity, namely
\be\label{rhoA-dual-prescription} Z_{QM}(L) = \int_{ \Gamma } [D\Gamma] \; Z(\Gamma)\,=\, \int \, \sqrt{g} d^2 x_\Gamma \; Z(x_\Gamma)\;~~~~~~~~\Gamma = B \cap \bar B ~~, \ee
where $Z_{QM}(L) $ is the partition function of the Quantum Mechanics system on the L side and $Z(\Gamma)$ is the gravitational partition function corresponding to the SS sector $\Gamma$ taken to be uniquely labeled by the tip position $x_\Gamma $.
This expression will be useful in our analysis, it follows from eq. \eqref{rhoA-osum} by taking trace, summing over the $\Gamma$-blocks. 
This expression for the partition function can be interpreted as follows:  the point $\Gamma$ in the bulk is undetermined \emph{a priori}, and so in principle, one should sum over all possibilities. The right hand side of this expression was used in the paper \cite{Witten2020b} for perturbative analysis. 

Note that the Hayward term is nothing but the Nambu-Goto action for the (pointlike) embedding field $x_\Gamma: \Gamma \to M$ and the corresponding eom is the condition of minimum for the Dilaton field $\Phi_\Gamma \equiv \Phi(x_\Gamma)$ \cite{Botta2020}. This encodes the "cosmic brane" of the model \cite{Dong2016} to compute R\'enyi entropies.

\subsection*{SS sectors as representations and Edge modes}

Gravity in 2d can be formulated as a gauge theory. The gauge group in principle is $SL(2,{\mathbb{R}})$, but for certain boundary conditions that imply conical singularities ($\theta\neq 2\pi$) this can be broken to a different residual symmetry group \cite{Mertens2019, Witten2020a, Mefford2020}.

In gauge theories, the SS sectors are representations, and the labels are given by the eigenvalues of the Casimir operators, so it can be expected that the area of the entangling surface ($\Phi_\Gamma$) be an observable associated to a Casimir of the symmetry group \cite{Takayanagi2019}.
In this sense, one of the achievements of the present approach is that the different SS sectors
 are weighted by probabilities that depend on the value of the field $\Phi_\Gamma$ on each splitting point $\Gamma$ as 
\be \label{rhoA-n-osum} \rho(L) = \bigoplus_{\Gamma}\; N \,e^{\frac{(2\pi - \theta )\Phi_\Gamma }{8 \pi G_N}}\; \rho_{JT} (\Gamma)= \bigoplus_{\Gamma}\; p_\Gamma\; \rho_{JT} (\Gamma)\;,\qquad\qquad p_\Gamma \equiv N \,e^{\frac{(2\pi - \theta )\Phi_\Gamma }{8 \pi G_N}}\;,
\ee
where $\theta$ is the opening angle on the corner of the geometry and $N$ is a normalization factor\footnote{The direct sum over $\Gamma$ sectors is formal yet, so this factor should be $N^{-1}=\int d\Gamma \, \mu(\Gamma) \,\,e^{\frac{(2\pi - \theta )\Phi_\Gamma}{8 \pi G_N}}\; $ where $\mu$ is the \emph{number} of entangling surfaces with area $\Phi_\Gamma$. }.
The following remarks are consequences of this result:
\begin{itemize}
\item[(a)] The SS sectors of \eqref{rhoA-osum} correspond to \emph{representations} of the symmetry group, so as for ordinary gauge theories, and the label $\Gamma$ is determined by the Casimir operators. 
\item[(b)] $\Phi_\Gamma$ are the eigenvalues of a Casimir operator that commutes with all the Algebra generators.
\end{itemize}
Therefore by virtue of (b), we note that the weight prefactors are in the center of the algebra, and one get a non trivial restriction to the gauge group ($SL(2,{\mathbb{R}})$ or residual): it has $\Phi_\Gamma$ as one of its Casimir operators.
In particular this is a Casimir of $SL(2,{\mathbb{R}})$ (e. g. see \cite{Jafferis2019}) and the unitary representation are infinite dimensional \cite{Kitaev2017}. 
It is worth pointing out here the possible presence of edge modes encoded in the Hayward term.
\begin{figure}
\begin{subfigure}{0.49\textwidth}\centering
\includegraphics[width=.9\linewidth] {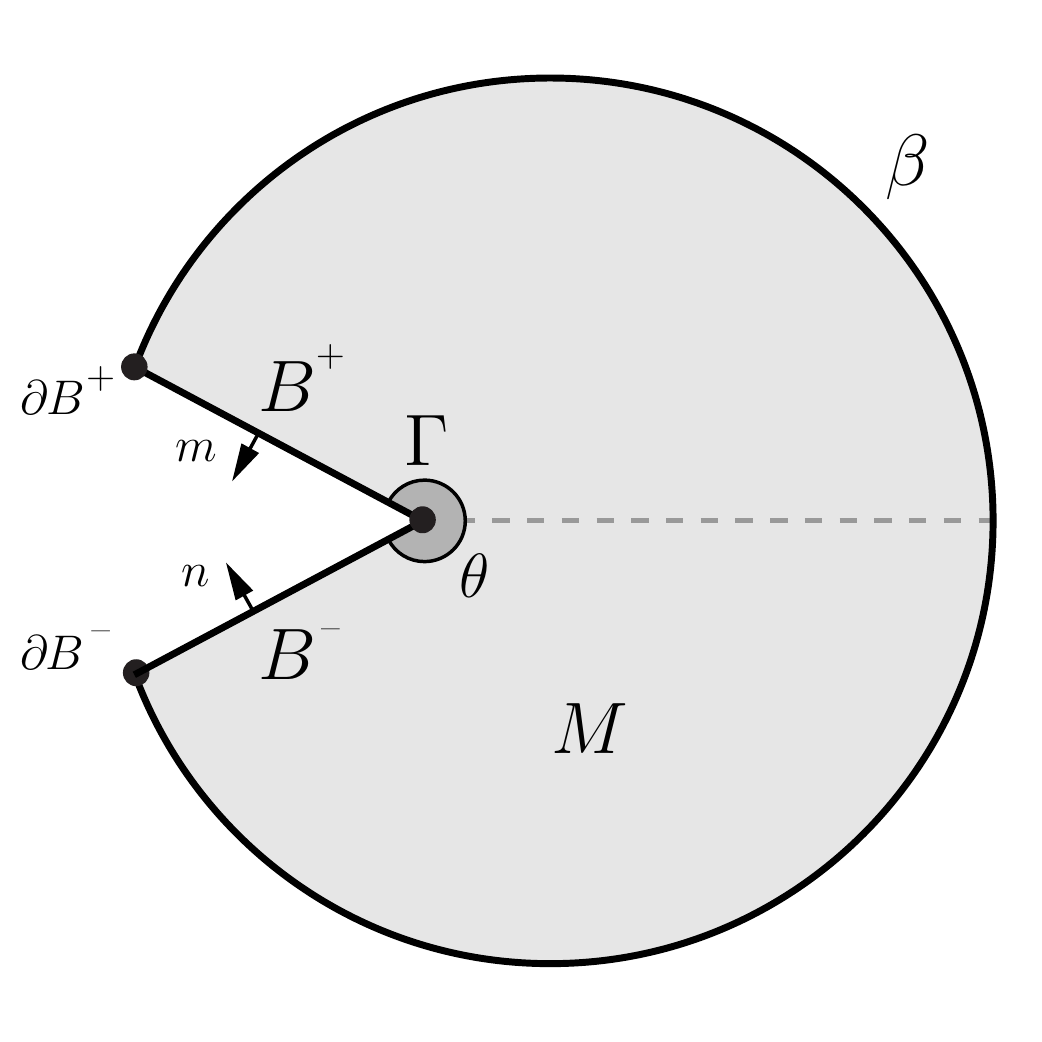}
\end{subfigure}
\begin{subfigure}{0.49\textwidth}\centering
\includegraphics[width=.9\linewidth] {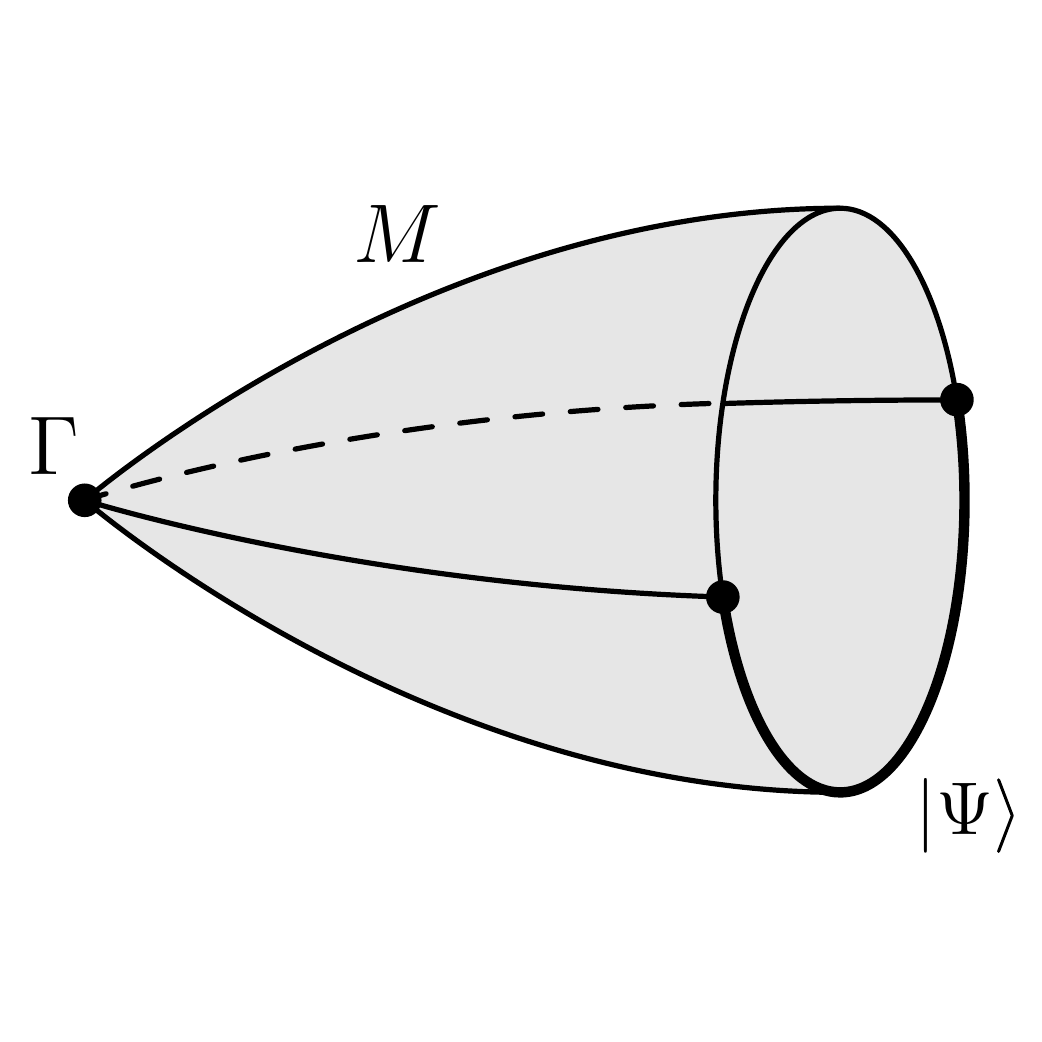}
\end{subfigure}
 \caption{ (a) A representation in the 2d plane of the computation of $\langle \phi^+|\rho|\phi^-\rangle$ is shown, where $\rho=\text{Tr}_{\bar B}|\Psi\rangle\langle\Psi|$. The $\phi^{\pm}$ define the field configurations in the branches. (b) The geometry associated to the partition function $Z[\Gamma]$ is shown. The computation is made by taking trace of the density matrix $\rho(\Gamma)$ within the $\Gamma$ SS sector.
The original state $|\Psi\rangle$ is represented by the lower semi-disk with a thicker line.}
\label{fig:Pacman}
\end{figure}

\subsection*{Edge modes}

The idea that in gravity the edge modes should be related to the \emph{corner} dynamics was pointed out first in \cite{Takayanagi2019}. 
In order to describe \emph{edge modes} in a gauge formulation of JT gravity, if the distribution $p_\Gamma$ is that given in are those of eq. \eqref{rhoA-n-osum}, the density matrix shall be given by
\be \label{rhoA-con edge} 
\rho(L) = \bigoplus_{\Gamma}\; p_\Gamma\; \rho_{JT} (\Gamma) \otimes \frac{I}{ \dim R_\Gamma},
\ee
where the last factor describes the edge modes \cite{Jafferis2019, Lin2018, Takayanagi2019}. The goal here is to estimate the dimensions of the representations $R_\Gamma$.
The entanglement entropy in this case writes
\be\label{entropy-conedge} S =  \sum_\Gamma \;- p_\Gamma \, \log p_\Gamma\; +\; p_\Gamma \, \log \dim R_\Gamma + \dots\;\;.\ee
This computation involves $\text{Tr} \;\rho$ associated to closing the Pacman into a geometry with conical deficit angle $\alpha = 2\pi-\theta$. The first term of this is the Shannon entropy that can be computed directly, the second one is related to the edge modes representations and $\ldots$ represent the terms associated to the average entropy of $\rho_{JT}$, which is distillable \cite{Trivedi2015,Lin2018-2}.  We shall concentrate in the terms that appear in this expression. Using \eqref{rhoA-con edge} we have
\be\label{entropy-conedge2} S =  \sum_\Gamma \; p_\Gamma \, \left( \frac{(\theta - 2\pi )\Phi_\Gamma }{8 \pi G_N}\; +\; \log \dim R_\Gamma \right)  
+ \dots\ee

On the other hand, the Ryu-Takayanagi formula for the (renormalized) entanglement entropy states 
\be\label{RTentropy} S  
= \frac{\langle \Phi_\Gamma \rangle}{4 G_N} \equiv \sum_\Gamma \; p_\Gamma \, \left( \frac{ \Phi_\Gamma }{4 G_N} \right)  \,.\ee
Therefore, comparing this with \eqref{entropy-conedge2} in the limit where the deficit angle of the conical geometry vanishes, we obtain the relation among expectation values in the state \eqref{rhoA-con edge}

\be \label{Lin} \langle \log \dim R_\Gamma \rangle \;\approx \;\frac{\langle \Phi_\Gamma \rangle}{4 G_N}\;, 
\ee 
consistent with previous results obtained from analysis of the measure of $SL(2,{\mathbb{R}})$ representations \cite{Lin2018}.

This argument even suggests a \emph{stronger} relation between the area/Dilaton eigenvalues and the dimension of the individual representations 
\be\label{conjecture-dimR} \log \dim R_\Gamma\; \approx \;\frac{ \Phi_\Gamma }{4 G_N}  \;\;.\ee
This expression would be quite interesting because
one can plug this back into \eqref{rhoA-con edge} and the factor $e^{\frac{ \Phi_\Gamma }{4 G_N}}$ cancels out with the same factor of $p_\Gamma$, then the weights of the representations becomes monotonically decreasing with the area: $p_\Gamma\sim \,e^{\frac{ - \theta \, \Phi_\Gamma }{8 \pi G_N}}\, $ for all $\theta>0$, as expected by the Ryu-Takayanagi statement. 

On the other hand, following \cite{Lin2018} we have considered here the case of a disk geometry, however for suitable boundary conditions we can have saddle geometries with conical singularity, which implies the generalization of \eqref{Lin} (or even more strongly \eqref{conjecture-dimR})
\be\label{dimR-generaltheta}
 \log \dim R_\Gamma = \frac{(2\pi + \alpha )\Phi_\Gamma }{8 \pi G_N}
\ee

The derivation of \eqref{Lin} in \cite{Lin2018} was based on the analysis of the Plancherel measure of the infinite-dimensional representations of $SL(2R)$. 
Although it can be expected that the symmetry changes because of the defect, this result is an evidence for it and \emph{quantifies} the deformation of the measure of the reps (linear in $\alpha$), or the symmetry group itself. 
Since in some models $\alpha$ is interpreted as the tension of a cosmic brane \cite{Dong2016}, this formula should be studied more in depth in higher dimensions and in connection with the cosmic brane dofs.

\subsection*{Fixed area sectors instead of  ``fixed area states''}

According to the construction in \cite{Dong2018}, the partition function clearly interprets as the modulus of a \emph{fixed area state}: $Z(\Gamma)\equiv \langle \Psi, \Gamma | \Psi, \Gamma \rangle $ for (nearly) eigenstates of the field $\Phi_\Gamma$, promoted to be an operator. However  from our perspective \cite{Botta2020}, summarized above, to fix the entangling (codim-2 surface) point $\Gamma$ in the bulk, corresponds to fixing the representation of the density matrix $\rho(\Gamma)$ of \eqref{rhoA-osum}: the SS sector, and the interpretation of this partition function is the trace of one specific block
\be Z(\Gamma) = \text{Tr}_\Gamma \, \rho(\Gamma). \ee
Even though that these two points of view can be quantitatively equivalent in the current JT theory, they are conceptually very different and the differences can drive to subtly different constructions. 

\section{The Hayward term in JT gravity}\label{HayenJT}

In this section we highlight the main pieces of the argument leading to the necessity of a Hayward term in the gravitational Dirichlet variational problem and compute its on shell action. In higher dimensional scenarios, where one works directly with the Einstein-Hilbert action, this problem has indeed been explored \cite{Takayanagi2019,Botta2020}, but we have not found this argument for the JT action, which due to the Dilaton requires some extra work. Other authors \cite{Harlow2018} have argued in favor of the presence of a similar contribution due to a topology argument, but this does not extend to the dynamical piece of the action and ultimately hides the fact that requiring a well posed variational Dirichlet problem is enough to completely fix the action and that the Hayward term is as necessary as the Gibbons-Hawking codim-1 term.

\subsection{Hayward term from a variational Dirichlet problem in gravity} 

We start from the JT bulk terms and we study it over the manifold $M$ shown in Fig. \ref{fig:Pacman}(a) representing a density matrix computation.  We can split the bulk contributions in topological and dynamical as
\begin{equation}\label{HJaffAction}
I_{JT}=I_{T} +I_{D} = \Phi_0  \int_M\!\! \sqrt{g} R  + \int_M\!\! \sqrt{g} \;\Phi\; (R-2\Lambda).
\end{equation}
Here $\Phi$ is the Dilaton field and $\Phi_0\gg1$ is a constant. 

Before moving on, a comment is due regarding the internal angle notation $\theta$. Notice that in Fig. \ref{fig:Pacman} we have called the internal angle $\theta$ rather than $\theta/2$ as in Fig. \ref{fig:H2}. This is because, in analogy with the TFD state's temperature $\beta$, we are taking the HH state to have internal angle $\theta/2$ and the associated density matrix (built by gluing two HH states along a subsystem) to have internal angle $\theta$. It should always be clear from the context which one is the correct angle to consider. This comment is especially relevant in the light of the Hayward term explicitly breaking the linearity of the EH action \cite{Hayward1993,Takayanagi2019,Botta2020} that was an important piece of the argument in \cite{Dong2016}, so one should always be aware of the object one is computing in our set-up.

\subsubsection*{Topological piece} 
 
By varying the Einstein-Hilbert action in 1+1 we get,
\begin{equation}
\delta\left( \int_M\!\! \sqrt{g} R\right) = \int_M\!\!  R  \delta \sqrt{g} + \int_M\!\! \sqrt{g} \delta R\nonumber\;.
\end{equation}
Using standard relations,
$$ \delta \sqrt{g} = -\frac 12 \sqrt{g}g_{\mu\nu}\delta g^{\mu\nu} 
\qquad\qquad 
\delta R = R_{\mu\nu}\delta g^{\mu\nu} + \nabla_\rho\left( g^{\sigma\nu}\delta\Gamma^\rho_{\nu\sigma}-g^{\sigma\rho}\delta\Gamma^\nu_{\nu\sigma} \right)\;,
$$
we get,
\begin{equation}
\int_M\!\! R  \delta \sqrt{g} = -\frac 12 \int_M\!\! \sqrt{g}\; R \; g_{\mu\nu}\delta g^{\mu\nu}\nonumber
\end{equation}
\begin{align}\label{Stokes}
\int_M\!\! \sqrt{g}  \delta R &= \int_M\!\! \sqrt{g} R_{\mu\nu}\delta g^{\mu\nu} + \int_M\!\! \sqrt{g} \nabla_\rho\left( g^{\sigma\nu}\delta\Gamma^\rho_{\nu\sigma}-g^{\sigma\rho}\delta\Gamma^\nu_{\nu\sigma} \right) \\
& = \int_M\!\! \sqrt{g} R_{\mu\nu}\delta g^{\mu\nu} + \int_{\partial;B^\pm}\!\! \sqrt{h} n_\rho\left( g^{\sigma\nu}\delta\Gamma^\rho_{\nu\sigma}-g^{\sigma\rho}\delta\Gamma^\nu_{\nu\sigma} \right)\nonumber
\end{align}
The bulk pieces provide the EH equations of motion, which in 1+1 is a geometrical identity
\begin{equation}
\int_M\!\! \sqrt{g} \left( R_{\mu\nu}-\frac 12 R \; g_{\mu\nu}\right) \delta g^{\mu\nu} = \int_M\!\! \sqrt{g} G_{\mu\nu} \delta g^{\mu\nu}=0\nonumber
\end{equation} 
This can be proven easily by reminding that $R^\rho_{\mu\sigma\nu} = R/2 (\delta^{\rho}_\sigma g_{\mu\nu}- \delta^{\rho}_\nu g_{\sigma\mu})$ is the most general Riemann tensor in 2d, thus by contracting one gets $R_{\mu\nu}=\frac R2 \; g_{\mu\nu}$.  
We are now left with the boundary terms
\begin{equation}\label{bound0}
\int_{\partial;B^\pm}\!\! \sqrt{h}  n_\rho\left( g^{\sigma\nu}\delta\Gamma^\rho_{\nu\sigma}-g^{\sigma\rho}\delta\Gamma^\nu_{\nu\sigma} \right)=\int_{\partial;B^\pm}\!\! \sqrt{h}  \left(n_\rho g^{\sigma\nu}\delta\Gamma^\rho_{\nu\sigma}-n^{\rho}\delta\Gamma^\nu_{\nu\rho} \right)
\end{equation}
where we have defined $n_\mu$ the external normal vector to each surface. In particular, we will also need to know the form of $\delta n_\mu$. We define this vector via a scalar function $\Upsilon(x^{\mu})=0$ 
such that $n_\mu= C \partial_\mu \Upsilon$ 
normalized such that $n^2=+1$. It is standard to consider a variation such that the $\Upsilon(x^{\mu})=0$ condition is unaffected by the variation. However, since the metric itself is modified, so does the normalization $C$
\begin{equation}
\delta n_\mu= \frac{\delta C}{C} n_\mu\;, \qquad\qquad \frac{\delta C}{C}=-\frac{1}{2} n_\alpha n_\beta \delta g^{\alpha\beta},\nonumber
\end{equation}
from where we see that $\delta n_\mu\propto n_\mu$ which we will use repeatedly below.
The expression \eqref{bound0} can be rewritten in terms of the variation of the normal vectors to each surface $n^\mu$ by using
\begin{equation}
\delta (\nabla_\sigma n_\nu) = \nabla_\sigma \delta n_\nu-(\delta \Gamma^\rho_{\nu\sigma})n_\rho \qquad\qquad \delta (\nabla_\mu n^\mu) = \nabla_\mu \delta n^\mu+(\delta \Gamma^\nu_{\nu\rho})n^\rho
\end{equation}
Considering the $B$ term for concreteness, we get
\begin{align}\label{bound1}
\int_{B}\!\! \sqrt{h}  \left(n_\rho g^{\sigma\nu}\delta\Gamma^\rho_{\nu\sigma}-n^{\rho}\delta\Gamma^\nu_{\nu\rho} \right)
&=\int_{B}\!\! \sqrt{h}  \left(\nabla^\nu \delta n_\nu-g^{\sigma\nu}\delta(\nabla_\sigma n_\nu)-\delta (\nabla_\mu n^\mu)+\nabla_\mu \delta n^\mu \right)\\
&=\int_{B}\!\! \sqrt{h}  \left(\nabla_\mu (\delta n^\mu +g^{\mu\nu}\delta  n_\nu)+(\delta g^{\sigma\nu})\nabla_\sigma n_\nu-2\delta K \right)\nonumber
\end{align}
where we used that the trace of the extrinsic curvature can be written as $K=\nabla_\mu n^\mu$. Now this is pretty close to what we are after. We have identified the origin of the Hayward term in the first term of the second line above. In order to use once again the Stokes theorem on it, one should first relate $\nabla_\mu \to D_a$, with $D_a$ being the covariant derivative compatible with the induced metric on $B$. This can be done by extracting the derivatives parallel to $n$, 
\begin{equation}\label{delta-u}
\delta u^\mu \equiv \delta n^\mu +g^{\mu\nu}\delta  n_\nu = n_\nu \delta g^{\mu\nu}  +2g^{\mu\nu}\delta  n_\nu,\nonumber
\end{equation}
\begin{equation}
\nabla_\mu \delta u^\mu = D_\mu \delta u^\mu - (n^\rho \nabla_\rho n_\mu)\delta u^\mu = D_\mu \delta u^\mu - (n^\rho \nabla_\rho n_\mu)n_\nu \delta g^{\mu\nu}\nonumber  
\end{equation}
where we have used that $\delta n_{\nu} \propto n_{\nu}$ so that $(n^\rho \nabla_\rho n^\nu) \delta n_{\nu} = 0$. We thus get
\begin{align}
\int_{B} \sqrt{h}  \left(n_\rho g^{\sigma\nu}\delta\Gamma^\rho_{\nu\sigma}-n^{\rho}\delta\Gamma^\nu_{\nu\rho} \right)
&=\int_{B}\sqrt{h}  \left(D_\mu \delta u^\mu + (\nabla_\mu n_\nu-n^\rho (\nabla_\rho n_\mu) n_\nu) \delta g^{\mu\nu} -2\delta K \right)\nonumber\\
&=\int_{B} \sqrt{h}  D_\mu \delta u^\mu + \int_{B} \sqrt{h}  K_{\mu\nu} \delta g^{\mu\nu} - 2\int_{B} \sqrt{h} \delta K\nonumber \\
&= \left[t_\mu  \delta u^\mu\right]_{\partial B;\Gamma} + \int_{B} \sqrt{h} ( K_{\mu\nu}-K h_{\mu\nu}) \delta h^{\mu\nu} - \delta\left(2\int_{B} \sqrt{h}  K \right)\nonumber\\
&= \left[t_\mu  \delta u^\mu\right]_{\partial B;\Gamma} - \delta\left(2\int_{B} \sqrt{h}  K \right),\nonumber
\end{align}
where the last term is the standard Gibbons-Hawking-York (GHY) boundary term for codim-1 boundaries in a manifold, in which the full $K_{\mu\nu}=\nabla_\mu n_\nu-n^\rho (\nabla_\rho n_\mu) n_\nu$ definition was used. We put $\delta g^{\mu\nu}=\delta h^{\mu\nu}$ when contracted with the extrinsic curvature and induced metric since all other components vanish by definition.

In the 1+1 dimensional set-up, the GHY term vanishes identically. This can be seen by considering that an extrinsic curvature in 1+1 is just a function and the induced metric is one dimensional. Thus, the extrinsic curvature tensor being a symmetric tensor on its indices must be of the form $K_{\mu\nu}=K h_{\mu\nu}$, regardless of the surface. We are left to explore the object in the first term. In it, we defined $t_a$ as the vector tangential to $B$ but normal to its boundaries. 
The first realization of such an object is that by definition $t^\mu n_\mu=0$, and $\delta n_{\nu}\propto n_{\nu}$, so that using \eqref{delta-u} one gets
\begin{equation}
t_\mu  \delta u^\mu = t_\mu  n_\nu \delta g^{\mu\nu}\nonumber.
\end{equation}
We thus find that the contribution we are after is related to the non diagonal element in the metric decomposition in terms of the $t$ and $n$ vectors. Furthermore, we find that this is not a full variation of any quantity, i.e. this isolated contribution cannot be regarded as a boundary term to the action. To do so, we must consider the similar contribution coming from other boundaries. Take for example the contributions coming from both $B^\pm$ at $\Gamma$, see Fig. \ref{fig:Pacman}(a)
\begin{equation}
(t_\mu  n_\nu -k_\mu  m_\nu )\delta g^{\mu\nu}.\nonumber
\end{equation}
These contributions require a definition of two basis of vectors $\{n,t\}$ coming from $B$ and $\{m,k\}$ coming from $\bar B$. Being in Euclidean signature, these bases must be related via a rotation. In particular,
\begin{equation}
m^{\mu}=-\cos\theta n^{\mu}+\sin\theta t^{\mu} \qquad\qquad n^{\mu}=-\cos\theta m^{\mu}-\sin\theta k^{\mu}\qquad\qquad -\cos\theta
=g^{\mu\nu}n_\mu m_\nu,\nonumber
\end{equation} 
from these relations one can put the $\Gamma$ contribution just in terms of the normal vectors as
\begin{align}
(t_\mu  n_\nu -k_\mu  m_\nu )\delta g^{\mu\nu}&=\frac{1}{\sin\theta}\left(m_\mu n_\nu+n_\mu m_\nu\right)\delta g^{\mu\nu}+\frac{\cos\theta}{\sin\theta}\left(m_\mu m_\nu+n_\mu n_\nu\right)\delta g^{\mu\nu}\nonumber\\
&=\frac{2}{\sin\theta}n_\mu m_\nu \delta g^{\mu\nu}+\frac{2}{\sin\theta}g^{\mu\nu} \delta(m_\nu  n_\mu)  
\end{align}
where we have used that $\delta n_\alpha = -\frac 12 n_\mu n_\nu \delta g^{\mu\nu} n_\alpha$. This can be seen to match the variation of the $\theta$ angle as defined via the normal $n,m$ vectors,
\begin{equation}
-2\delta(\cos\theta)=2\sin\theta\;\delta\theta = 2n_\mu m_\nu \delta g^{\mu\nu}+2g^{\mu\nu} \delta(n_\mu m_\nu)\;\nonumber.
\end{equation}
One gets a similar contribution from all corners.
We have finally proven that
\begin{equation}
\delta\left( \int_M\!\! \sqrt{g} R + 2\int_{\partial;B^\pm}\!\! \sqrt{h} K\right) =  
2\delta\theta_\Gamma+2\delta\theta_{\partial B}+2\delta\theta_{\partial \bar B}\nonumber
\end{equation}
The codim-2 terms we got from the topological pieces (upon considering the complete manifold $M$) are indeed full variations and can be removed via boundary terms. 
This is interpreted as these corners not adding any extra dofs in an EH action in 1+1, i.e. the action is still only topological. 
The action that provides a well posed problem on this regard is 
\begin{equation}
I_T+2\Phi_0(\theta_\Gamma-2\pi)+2\Phi_0( \theta_{\partial B^+}-\pi/2) + 2\Phi_0(\theta_{\partial  B^-}-\pi/2)
\end{equation}
where the $2\pi$ factor in $(\theta_\Gamma-2\pi)$ is fixed so that we do not have a Hayward term when there is no deficit angle in the geometry corresponding to the density matrix. On the other hand, since $\theta_{\partial B^\pm}$ belong to a codim-2 corner that represents a timelike-spacelike corner in real-time signature, their contributions are expected to vanish for $\theta_{\partial B^\pm}=\pi/2$.

\subsubsection*{Dynamical piece:} 
 
The treatment on the dynamical piece of the action follows along the lines of the topological piece but one should be careful with the extra terms appearing due to the Dilaton. We start from
\begin{equation}
 I_{D} = \int_{M}\!\! \sqrt{g} \;\Phi\; (R-2\Lambda)\nonumber 
\end{equation}
whose variation is
\begin{equation}
 \delta I_{D} = \int_{M}\!\! \sqrt{g}\; (R-2\Lambda)\;\delta \Phi + \int_{M}\!\! \sqrt{g}\; \Phi\;\delta(R-2\Lambda)+ \int_{M}\!\! \delta(\sqrt{g})\; \Phi\;(R-2\Lambda) \;.\nonumber
\end{equation}
The first term above will provide eoms for the metric that fix $R=2\Lambda$ even off shell, since the Dilaton can be integrated out exactly in the path integral. The second and third terms lead to
$$\int_{M}\!\! \delta(\sqrt{g})\; \Phi\;(R-2\Lambda) =\frac 12 \int \sqrt{g} \;\Phi\; (R-2\Lambda)g^{\mu\nu}\delta g_{\mu\nu}, $$
$$\int_{M}\!\! \sqrt{g}\; \Phi\;\delta R = \int_M\!\! \sqrt{g}\; \Phi\; R_{\mu\nu}\delta g^{\mu\nu} + \int_M\!\! \sqrt{g}\; \Phi\; \nabla^\mu\left( \nabla^\nu\delta g_{\mu\nu}-g^{\nu\rho}\nabla_\mu\delta g_{\nu\rho}\right),$$
where the second term should be manipulated as
\begin{align*}
 \Phi\; \nabla^\mu\left( \nabla^\nu\delta g_{\mu\nu}-g^{\nu\rho}\nabla_\mu\delta g_{\nu\rho}\right)
&= \nabla^\mu\left[ \Phi\;  \left( \nabla^\nu\delta g_{\mu\nu}-g^{\nu\rho}\nabla_\mu\delta g_{\nu\rho}\right)\right] - \left( \nabla^\mu \Phi \right)\;  \left( \nabla^\nu\delta g_{\mu\nu}-g^{\nu\rho}\nabla_\mu\delta g_{\nu\rho}\right)\;.
\end{align*} 
Further manipulations on the second term above lead to the Dilaton eoms plus some boundary terms,
$$\int \sqrt{g}\left[ -R^{\mu\nu}\Phi+ \nabla^\mu\nabla^\nu\Phi - g^{\mu\nu}\nabla^2\Phi \right] \delta g_{\mu\nu}+\int_{\partial; B; \bar B}\!\! \sqrt{h}\; n_\mu \nabla^\mu \Phi \;h^{\nu\rho}\; \delta h_{\nu\rho}+
\int_{\partial;B^\pm}\!\! \sqrt{h} \; \delta u^\mu \;D_\mu \Phi$$
Recalling that in 1+1 we have $R^{\mu\nu} = R/2 g^{\mu\nu} = - g^{\mu\nu}$ we get,
\begin{equation}
    g_{\mu\nu}\left[ g^{\mu\nu}\Phi+ \nabla^\mu\nabla^\nu\Phi - g^{\mu\nu}\nabla^2\Phi \right]= 2\Phi- \nabla^2\Phi =0
    \qquad\Rightarrow\qquad
(\nabla^\mu\nabla^\nu - g^{\mu\nu})\Phi =0\nonumber
\end{equation}
Our main interest arises in the boundary terms rising from
\begin{align}\label{dinbc0}
 \int_M\!\! \sqrt{g}\; \nabla^\mu\left[ \Phi\;  \left( \nabla^\nu\delta g_{\mu\nu}-g^{\nu\rho}\nabla_\mu\delta g_{\nu\rho}\right)\right]=\int_{\partial; B^\pm}\!\! \sqrt{h}\; n^\mu\left[ \Phi\;  \left( \nabla^\nu\delta g_{\mu\nu}-g^{\nu\rho}\nabla_\mu\delta g_{\nu\rho}\right)\right]\nonumber
 \end{align}
From our work on the topological term, this boundary contribution can be written as
\begin{align}
 \int_M\!\! \sqrt{g}\; \nabla^\mu\left[ \Phi\;  \left( \nabla^\nu\delta g_{\mu\nu}-g^{\nu\rho}\nabla_\mu\delta g_{\nu\rho}\right)\right]&= \int_{\partial;B^\pm}\!\! \sqrt{h} \;\Phi\; D_\mu \delta u^\mu - 2  \int_{\partial;B^\pm}\!\! \;\Phi\;\delta\left(\sqrt{h} K \right) + \int_{\partial;B^\pm}\!\! \sqrt{h} \;\Phi\; ( K h_{\mu\nu}-K_{\mu\nu})\delta h^{\mu\nu}\nonumber\\
&= \int_{\partial;B^\pm}\!\! \sqrt{h} \;\Phi\; D_\mu \delta u^\mu -  \delta\left(2\int_{\partial;B^\pm}\!\! \;\Phi\;\sqrt{h} K \right)+ 2 \int_{\partial;B^\pm}\!\! \sqrt{h} K \;\delta\Phi \;\nonumber,
\end{align}
where recall that $K h_{\mu\nu}=K_{\mu\nu}$ is a geometric identity in 1+1. The second and third terms above can no longer be manipulated.
The first term is the one that contains the codim-2 contributions, i.e.
\begin{align}
\int_{\partial;B^\pm}\!\! \sqrt{h} \;\Phi\; D_\mu  \delta u^\mu = \int_{\partial;B^\pm}\!\! \sqrt{h} \;D_\mu (\Phi\;  \delta u^\mu) - \int_{\partial;B^\pm}\!\! \sqrt{h} \; \delta u^\mu \;D_\mu \Phi\;.
\end{align}
The second term cancels a contribution that appeared earlier, whilst the first term finally yields, again by virtue of our $I_T$ piece analysis,
\begin{equation}
\int_{\partial;B^\pm}\!\! \sqrt{h} \;D_\mu (\Phi\;  \delta u^\mu)=   2 \; \left( \Phi_\Gamma \delta\theta_\Gamma + \Phi_{\partial B^+} \delta\theta_{\partial B^+} + \Phi_{\partial  B^-} \delta\theta_{\partial B^-}\right)\nonumber
\end{equation}
We arrive at
\begin{align}\label{eqforAppB}
\delta I_D +\delta\left( 2\int_{\partial;B^\pm}\!\! \sqrt{h} \Phi K\right)&=  \delta\left( \int_M\!\! \sqrt{g} \Phi(R-2\Lambda) + 2\int_{\partial;B^\pm}\!\! \sqrt{h} \Phi K\right)\nonumber \\ & =  
 2 \int_{\partial;B^\pm}\!\! \sqrt{h} K \;\delta\Phi +\int_{\partial; B^\pm}\!\! \sqrt{h}\; n_\mu \nabla^\mu \Phi \;h^{\nu\rho}\; \delta h_{\nu\rho} + 
 2 \left( \Phi_\Gamma \delta\theta_\Gamma + \Phi_{\partial B^+}\delta \theta_{\partial B^+} + \Phi_{\partial  B^-}\delta \theta_{\partial  B^-}\right)\nonumber\;.
\end{align}
The first two terms mandate the possible boundary conditions to impose at each boundary, either Dirichlet, $\delta h_{\nu\rho}=\delta \Phi=0$, or Neumann conditions,  $K=0$ and $n_\mu \nabla^\mu \Phi=0$. The last terms are not a full variation and are a signal of the existence of extra dofs at these points. From a variational problem viewpoint, one should further impose either $\delta \theta=0$ or $\Phi=0$ at the corners. Since we are interested in a Dirichlet problem $\delta h_{\nu\rho}=\delta \Phi=0$,
we see that $\delta \Phi=0$ for $\partial;B^\pm$ in turn induces $\delta \Phi=0$ also on the corners. The natural dynamical action for a Dirichlet problem on our Pacman geometry is thus
\begin{align}
 I_D + 2\int_{\partial;B^\pm}\!\! \sqrt{h} \;\Phi\; K 
  + 
 2 \; \left( \Phi_\Gamma (\theta_\Gamma-2\pi) + \Phi_{\partial B}(\theta_{\partial B}-\pi/2) + \Phi_{\partial \bar B}( \theta_{\partial \bar B}-\pi/2)\right)
\end{align}
which can be seen to be completely defined by $\delta h_{\nu\rho}=\delta \Phi=0$ including the corners.

\subsubsection*{Summary}

We started from the JT bulk terms and explored manifolds with codim-2 corners in the geometry, i.e. Pacman manifolds. We showed that in order to define a well posed Dirichlet problem from this metric one should consider Hayward terms. We thus now define the action we will be working with, which is
\begin{align}
I_{JTH} &\equiv I_T + I_D +2\int_{\partial;B;\bar B}\!\! \sqrt{h}( \Phi_0+\Phi) K -2\int_{\partial}\!\! \sqrt{h}\Phi \nonumber\\
&\qquad- 2(\Phi_0+\Phi_\Gamma) (\theta_\Gamma-2\pi) -2 (\Phi_0+\Phi_{\partial B^+})(\theta_{\partial B^+}-\pi/2) -2 (\Phi_0+\Phi_{\partial B^+})(\theta_{\partial B^-}-\pi/2)\label{JTH} 
\end{align}
Notice that an extra codim-1 term has been added which does not arise from our variational problem. This term is required rather by demanding the on shell action to be finite via holographic renormalization and it is only required at asymptotic boundaries \cite{Skenderis2002}.

\subsection{On-shell Solution} \label{onsolution}

Here we write the classical on shell action to the problem of JT gravity in presence of the Hayward term. We start from action \eqref{JTH} and
integrate out $\Phi$ fixing $R+2=0$. Notice crucially that the Hayward term is a boundary term and as such it does not modify the eoms outside of its boundary conditions.
The spacetime then is fixed to be AdS$_2$,
\be
ds^2=(r^2-1)d\tau^2+\frac{dr^2}{r^2-1},\qquad\qquad 0\leq\tau\leq\theta. \label{metricAdS2}
\ee
Here $\theta$ is the opening angle at the tip, the place where the conical defect will arise if one completes the space with the point. For $\theta=2\pi$ there is no conical defect. By considering the $B^\pm$ surfaces to be that of constant $\tau$, one can see that all codim-1 terms in the action involving these surfaces vanish. Even if one chooses different $B^\pm$ surfaces, these codim-1 contributions should cancel upon gluing in building the partition function, as we will do in the Sec. \ref{sources}. For now, we will consider the former fixed $\tau$ choice for simplicity.

The identification in the coordinate $\tau$ breaks the $SL(2,{\mathbb{R}})$ symmetry of the hyperbolic plane and just a diagonal $U(1)$ remains, see App. F of \cite{Mefford2020}. To deal with the boundary dofs we need to fix the boundary proper length, which will be done cutting a piece of AdS$_2$ to a nearly AdS$_2$ space by using a regulator $\epsilon$, setting
\begin{equation}\label{boundaryfields}
    \Phi_{bdy}=\frac{\phi_\partial(u)}{\epsilon},\qquad\qquad ds^2_{bdy}=\frac{du}{\epsilon},
\end{equation}
which implies that the proper length of the boundary is $L=\int_0^\beta ds=\beta/\epsilon$, where the time on the boundary curve runs in $u\in [0,\beta)$. At the end one has to take $\epsilon\rightarrow 0$.
Now, as is usual in JT gravity we will label the boundary curve by some parameter $u$ which has the range $0\leq u\leq \beta$. So, writing the line element \eqref{metricAdS2} as function of $u$ it can be seen that the remaining action (to leading order in $\epsilon$) is the one of a Schwarzian theory plus the Hayward contribution

\be
I_{JTH}= -\frac{\phi_b}{8\pi G_N}\int_0^\beta du\,Sch\left[\tan\frac{\tau(u)}{2},u\right]+\frac{1}{8 \pi G_N} (\Phi_0 + \Phi_\Gamma)(\theta - 2\pi),\nonumber\label{JTH2}
\ee
where we used a constant value $\phi_\partial(u)=\phi_b$ on the boundary, see App. \ref{schwarz} for the derivation of the first term in this coordinate system. Notice also that the Hayward terms arising from $\partial B^\pm$ are absent in the expression above. In App. \ref{ADM} we show via an ADM study that a gauge fixing allows to disregard them in our analysis unlike the Hayward term at $\Gamma$ that holds physical information. 

The Schwarzian theory is an action for a scalar field $\tau: S^1 \to S^1$ whose respective circumferences are $\beta, \theta$.
Given that the (Euclidean time) variable $u\in[0, \beta]$ parameterizes the circle, the field configurations  $\tau(u)$ are interpreted as different reparameterizations.  As boundary condition one shall demand that the periodicity $u\to u+ \beta$ is mapped to the new interval as $\tau(u + \beta) = \tau(u) + \theta$, or in other words, the field $\tau(u)$ can be thought of as a reparameterization map from a circle of length $\beta$ to a circle of length $\theta$. In cases where $\beta=2\pi n\;$, $n\in N$ we will call $n$ a winding number. In JT gravity these solutions are not stable \cite{Maldacena2016JT} but their analysis will be useful when we study the replica trick.

To proceed with the computation of the on-shell action we must say the functional form of the time coordinate. We will study a dominant solution which solves  the equation of motion of the Schwarzian theory
\be
\tau(u)=\frac{\theta}{\beta}u.\label{dominant}
\ee
Then, performing the integration in $u$ we arrive to 
\be
I_{JTH}=-\frac{\phi_b}{16\pi G_N}\frac{\theta^2}{\beta}+\frac{1}{8 \pi G_N} (\Phi_0 + \Phi_\Gamma)(\theta - 2\pi).\label{JTHonshell}
\ee
This of course can be re-written in terms of the deficit angle $\alpha=2\pi-\theta$ which resembles the conical defect on-shell solution of \cite{Mertens2019, Witten2020a, Witten2020b}.

\section{Sources and JT geometries with conical defects}\label{sources}

The goal of the current chapter is to highlight that there are at least two ways of producing conical defects in JT gravity. The commonly used is by inserting pointlike sources in the bulk as in \cite{ Witten2020b}. In our approach they will appear because of the boundary contributions due to wedge shaped geometries. In this section we will see a general analysis that capture both possibilities, but the second one has some advantages to compute R\'enyi entropies in holographic scenarios and to prove the JLMS proposal. In particular, we are going to conclude that the solutions of both formulations are equal but the on-shell actions differ by a contribution given precisely by the Hayward term.

\subsection{Pointlike sources in the bulk}

JT gravity is a consistent theory for the 2d spacetime geometry $(M,g_{\mu\nu})$ as well as for the Dilaton field $\Phi(x)$. The bulk defects that have been more studied in the literature are actually formulated as pointlike \emph{sources} for the Dilaton, generally introduced  through a coupling term $I_\alpha = 2 \int_M d^2x \, \alpha (x) \Phi(x)$ in the action \cite{Mertens2019, Witten2020a, Witten2020b}. Therefore, the total action is linear in $\Phi(x)$ and the corresponding equation of motion constraining the geometry is 
 \be \label{einsteineq-source}
 R(x) +2 = 2 \alpha(x) \qquad x \in Int[M]\;,
 \ee 
which has to be supplemented with the condition for the Dilaton $[\nabla_\mu \nabla_\nu -g_{\mu \nu} ]\Phi(x)=0 $. 
In particular, if we set a pointlike source  $\alpha(x) = \alpha \;\delta^2(x-x_0)$  at the point $x_0 \in M $, we will obtain an equation with source for the geometry $R+2= 2 \alpha\, \delta^2(x-x_0)$ whose solution is a conifold with angular deficit given by $2\pi-\theta= \alpha$.
This is the reason why we may refer to this configuration as \emph{bulk defect}.

In fact, the path integral of JT  gravity, including a source term is, recovering the $16\pi G$ factors,
  \be\label{ZJT-source}
 Z_{JT}[\phi_\partial, \beta, \alpha]\equiv \int\left[D\Phi(x)\right]_M [Dg(x)]_M \, e^{\frac{1}{16 \pi G_N} \int_{M} \; \sqrt{g} (R(x) + 2)\Phi(x)\,+ \,\frac{1}{8 \pi G_N}\, \int_{M} \; \sqrt{g} \alpha(x)\,\Phi(x) dx + \,\frac{1}{8 \pi G_N} \int_{\partial M} \sqrt{h} \Phi (K-1)} .
  \ee
Notice that we have ommited the topological term contribution containing $\Phi_0$ for it plays no fundamental role in the discussion.
This theory agrees with the action proposed in \cite{Almheiri2019, Ellerin2021} as alternative to the method of \cite{Dong2016} to evaluate $n$-th refined R\'enyi entropy for $\alpha \equiv 1-1/n$, since it describes a self graviting ''0-brane" object at $x_0$ with tension $T_n\propto\frac{n-1}{n}$.
The field $\Phi(x)$ can be exactly integrated out from the full path integral, resulting the \emph{off-shell} condition \eqref{einsteineq-source}. Plugging this into \eqref{ZJT-source} we obtain just an effective boundary theory
\be\label{ZJT-source-boundary}
 Z_{JT}[\phi_\partial, \beta, \alpha]\equiv \int  [Dg(x)]_{M(\alpha)} e^{ \,\frac{1}{8 \pi G_N} \int_{\partial M} \sqrt{h} \phi_\partial (K-1)} .
  \ee
where the measure $[Dg(x)]_M$ must be substituted by a sum over geometries satisfying \eqref{einsteineq-source}: $[Dg(x)]_{M(\alpha)}= [Dg(x)]_M \, \delta(R+2 - \alpha)$.
This partition function can be expressed as a Schwarzian theory for a boundary dof, that can be exactly valued \cite{Mertens2019}
   \be\label{ZJT-valued}
Z_{JT}[\phi_\partial, \beta, \alpha] = \left(-\frac{C}{\beta}\right)^{1/2}\; e^{- C\, \frac{(2\pi - \alpha)^2}{\beta} },\ee
where $C<0$ only depends on the boundary values $\phi_\partial$, that according to \eqref{boundaryfields} is given by $C =-\frac{\phi_b}{16\pi G_N}$.

We would like to end this section by stressing that these type of sources in the Euclidean geometry, 
are hard of arguing by only using holographic ingredients if they are to come from states built via a path integral. In our mind, the two dimensional Euclidean geometry arises in the holographic context, because of computations involving states of the dual QFT as the Hartle Hawking wave functional. So the type of pointlike sources described here are hard of arguing in such contexts. This is our main motivation to propose the set up and analysis below.

\subsection{Defects from boundary corners and Hayward contribution}

In this section we elaborate more on defects whose origin is a corner \emph{on the boundary}, such that the equation of motion is $$R+2=0 \quad \forall x\in Int[M]$$ in place of \eqref{einsteineq-source}. We refer to as \emph{boundary defects} or simply corners, and we  will show differences with the description of the bulk ones, as well as implications on certain computations such as R\'enyi entropies.

Remind from arguments of Sec. \ref{bipartite} that the reduced state of the boundary quantum mechanical
system can be decomposed in a direct sum on different gravitational subsystems. So 
the \emph{gravitational} reduced density matrix associated to the ``left'' of the point $x_0$ (the region $B$ in Fig \ref{fig:H2}(b))
in the bulk is 
 \be \rho (B, \beta) \equiv \text{Tr}_{{\cal H}_{\bar B}} \,|\Psi \rangle \langle \Psi |
= \sum_{\phi_{\bar B}} \,  \langle\phi_{\bar B}|\,\Psi \rangle\, \,\langle \Psi \,| \phi_{\bar B}\rangle \, .\nonumber
\ee
We are schematically considering a \emph{configuration basis} $\phi(B)\equiv (h|_B , \Phi|_B)$ on the spacelike interval $B$.
Defining two arbitrary field configurations $\phi^\pm \equiv \phi(B^\pm) = \phi(\pm i \beta/2)$ \footnote{In particular in standard JT gravity discussed in the previous subsection, this propagator/path integral involves only two entangled dof's associated to the two asymptotic modes  $\psi(x_L, x_R) \in {\cal H}_L \otimes {\cal H}_R$, thus, the partial trace is: $\rho(x^-_L, x^+_L) \equiv Tr_R \,\psi \,\psi^\dagger\, = \int dx_R \, \delta(x_R-y_R)\,\psi(x^-_L, x_R) \psi^\dagger(y_R , x^+_L)$. In Sec. \ref{livingontheedge} we argue that an edge mode associated to $\Gamma$ appears as an extra dof} on two copies (or branches) of the surface $B$, denoted as $B^\pm$, which intersect in a point $\Gamma = B^+ \cap B^-$, one can express its matrix elements as the product of euclidean evolution operators \cite{Botta2018, Botta2019}\footnote{The representation of the pure states in terms of evolution operators is convenient and more illuminating for the computations involving the replica method and was used in the past to compute the modular Hamiltonian for excited states in holography \cite{Arias2020}.}
\be\label{rho-psipsi}
\langle \phi^+ | \rho (B, \theta) |\phi^-\rangle = \sum_{\phi_{\bar B}}
\langle \phi^+_{}| \,U\left(-i\beta/2, 0\right)\,  |\phi_{\bar B}\rangle\langle\phi_{\bar B}| \,U\left(0, i\beta/2)\right)\,| \phi^-_{}\rangle = \langle \phi^+ | U\left(-i\beta/2 , i\beta/2 \right) |\phi^-\rangle,
\ee
where we have used the completeness of the configuration basis $I_{\bar B}\equiv\int {\cal D}\phi_{\bar B}|\phi_{\bar B}\rangle\langle\phi_{\bar B}|$ on ${\cal H}_{\bar B}$. This is well defined as a path integral, 
and one can compute this in the large $N$ approximation:

\be\label{rhoB} \langle \phi^+ | \rho (B, \theta) |\phi^-\rangle = \int^{}_{\phi^\pm}\left[D\Phi(x)\right]\, [Dg(x)] e^{-I_{JT}[\Phi, g, M_P]}
\approx \, e^{-I_{bulk}[\phi, M_P]} \, e^{-I_{bdy}[\phi^\pm ] + \frac{(2\pi - \theta) }{8\pi G} \Phi_\Gamma}\,,
\ee
where, by virtue of the saddle point approximation, we evaluated the action in a classical solution $M_P = M^- \cup M^+$ smoothly glued on the surface ${\bar B}$, whose boundaries are the branches (curves) $B^-$ and $B^+$ (see Fig \ref{fig:Pacman}a). 
This saddle manifold is clearly a Pacman geometry (Fig 2b) characterized by a corner with an opening angle $\theta$. In principle the boundary data $\phi^\pm$ that label the matrix elements (so as the asymptotic value $\phi_\partial$ characterizing the state), can backreact but the local metric is always $AdS_2$.

Remarkably, in this context the point $\Gamma$ \emph{belongs to the boundary}, and therefore, differently from \eqref{einsteineq-source}, the \emph{saddle} bulk geometry do not receive contributions from the Hayward term and so we always have 
\be\label{einstein-homog} R(x) + 2 = 0 , \;\; \forall x\in M_P \ee
The geometric counterpart (associated saddle) of $ \text{Tr} \rho$ is gluing together the intervals $B^\pm$, so $M_P$ becomes a conifold $M$ with deficit angle $2\pi - \theta$, which implies that the integral of curvature on $M$ is \emph{distributionally} consistent with the equation \eqref{einsteineq-source}: $\int_M d^2x \sqrt{g} \,(R+2) \,\Phi = 2(2\pi - \theta) \Phi(x_0) $. The important consequence of this construction is that the corner of $M_P$ becomes the tip of the cone $M(2\pi - \theta)$, see Fig. \ref{fig:Pacman}. 

In fact, defining the partition function as the trace of \eqref{rhoB} by taking $\phi^+=\phi^-$ on the (open) intervals $B^\pm$, which one sums over, results the JT partition function with a remaining Hayward term valued on $\Gamma$, which belongs to the boundary
\be\label{ZJT-bdy-defect}
 Z_{JT}[\phi_\partial, \beta, \alpha_\Gamma]\equiv 
 \int\left[D\Phi(x)\right]_{M_0} [Dg(x)]_{M_0}  e^{-\frac{1}{16 \pi G_N} \int_{M_0} \; \sqrt{g} (R(x) + 2)\Phi(x)\,+\, \int_{M} \; \sqrt{g} \alpha_\Gamma(x)\,\Phi(x) dx - \,\frac{1}{8 \pi G_N} \int_{\partial M} \sqrt{h} \Phi (K-1)},
  \ee
where $M_0 \equiv M - \Gamma$, and 
\be \alpha_\Gamma(x) \equiv \frac{(2\pi - \theta) }{8\pi G} \,\delta(x-x_0).\nonumber
\ee
In other words, the (conical) geometries on which we should sum over, do not include the tip point which must be considered as a part of the interior boundary. This partition function comes from a pure state that in the Hartle-Hawking formalism, can be described  as a path integral of Euclidean JT gravity without \emph{any defect} or source. But it emerges by taking the trace of a state built from these wave functionals in the bulk theory.

Let us see that this subtle conceptual difference with bulk sources reviewed above have important quantitative consequences, e.g. on the derivations of the holographic prescriptions to compute the R\'enyi entropies, as well as many other aspects related to holography and entanglement.

\subsubsection{Fixed Area \emph{Sectors} and gravity ensembles/partition functions}\label{fixedareaZ}

In JT gravity, the standard computation of  the path integral on a disc with (or without) source at $x_0$ (eq. \eqref{ZJT-source}) involves a measure for the Dilaton that can be writen in a factorized form
\be\label{Dphi} \int\, [D \Phi(x)]_{x\in M} = \int^{+\infty}_{-\infty}[d \Phi(x_0)]\,\int [D \Phi(x)]_{x\in M_0}\;,\qquad\qquad M_0\equiv M-\{x_0\}\;.\ee
The last path integral factor denotes a sum over Dilaton configurations on the manifold $M_0$. 
The same factorization of the measure can be formally expressed for the space of 2d-Euclidean metrics.
It is illuminating to consider the JT path integral fixing $\Phi_\Gamma$ \emph{a priori}, or integrating it out.
   
In the present context it is subtly different: the path integral \eqref{ZJT-bdy-defect} already supposes $\Phi_\Gamma$ fixed from the beginning (i),  
but regarding \eqref{Dphi}, one can also consider the integration of it (ii). Let us consider both possibilities in order to interpret and compare results in different saddle geometries.

We have these two cases:
\begin{itemize}
 \item[(i)]Fix the boundary data $\Phi(x_0)$, and do not integrate it out. Because in JT gravity the \emph{area} of a codim-2 surface $\Gamma$ in JT gravity is given by $\Phi(x_0)$, by fixing this one obtains a clear realization of the so-called \emph{fixed area states} \cite{Dong2018} in the JT laboratory.
We will show that this gives place to the \emph{flat spectrum} of the R\'enyi entropies.

\item[(ii)] Consider the full integration of this field expressed in \eqref{Dphi}. This realizes the integration on all the areas giving place to the Maldacena-Lewkowycz-Dong (MLD) smooth geometries \cite{Dong2016, Aitor2013} that will be useful to compute the refined R\'enyi entropies, which will have a non-trivial spectrum.
\end{itemize}
Moreover, since the relation between both possibilities lies on integrating out $\Phi(x_0)$, or not, in the partition function, the conjectured interpretation of both types of saddle as different \emph{ensembles} is automatically proved in the present JT context.
In the following we will study both cases in detail.

  \subsubsection*{Analysis of (i)}
 
  By fixing the value of the observable $\Phi(x_0)$, interpreted as the area in JT gravity, the partition function \eqref{ZJT-bdy-defect} expresses the trace of \emph{fixed area states} $\rho[\Phi(x_0)]$
 \begin{align}\label{ZC-fa-M0}
 Z[\phi_\partial, \beta, \theta, \Phi(x_0) ]&\equiv \text{Tr} \rho[\Phi(x_0)]\\
 &=
 \int\left[D\Phi(x)\right]_{M_0} \, \left[Dg(x)\right]_{M_0} \; C[\theta,\Phi(x_0)] \;  e^{-\frac{1}{16 \pi G_N} \int_{M_0} \; \sqrt{g}\Phi (R+2)\, - \,\frac{1}{8 \pi G_N} \int_{\partial M} \sqrt{h} \phi_\partial (K-1)},
 \end{align}
  where we see that the factor
  \be \label{CJT}
 P[\theta,\Phi(x_0)]\equiv \,e^{ \;  \,\frac{(2\pi - \theta(x_0))\,\Phi(x_0) }{8 \pi G_N}}\equiv \,e^{- \;  \,\frac{ \theta(x_0)\,\Phi(x_0) }{8 \pi G_N}} \;D[\Phi(x_0)] \,\,.\ee
  captures the tip/corner local features and can be viewed by its own as a operator inserted in a standard JT theory
\be\label{}
 Z[\phi_\partial, \beta, \theta, \Phi(x_0) ]\equiv \text{Tr}\,\left( P[\theta,\Phi(x_0)]\,\,\, \rho_{JT} \right)\;  \nonumber.
  \ee
It is worth emphasizing that we are making contact with the so called \emph{defect operator} proposed in \cite{Jafferis2019} from totally different  arguments. The referred defect operator consists of the non additive part of this operator, and independent on $\theta$
  \be D[\Phi(x_0)] = \,e^{ \;  \,\frac{\,\Phi(x_0) }{4 G_N}}\nonumber
 ~~~~\Rightarrow ~~P = D^{-\theta/2\pi +1}.
 \ee
By integrating out the fields $\Phi(x\in M_0) $ in \eqref{ZC-fa-M0}, we can eliminate the bulk contribution and obtain the remarkable expression
  \be\label{ZJTH-fixedA}
 Z[\phi_\partial, \beta, \theta, \Phi(x_0) ]= \; 
\,\int \left[Dg(x)\right]_{M(2\pi-\theta)} \;\,e^{ \;  \,\frac{(2\pi - \theta(x_0))\,\Phi(x_0) }{8 \pi G_N}}\;  e^{- \,\frac{1}{8 \pi G_N} \int_{\partial M} \sqrt{h} \phi_\partial (K-1)}.
\ee
 The measure $\left[Dg(x)\right]_{M_0}$ above became restricted to a sum over all the 2d geometries such that $R+2=0,\;\; \forall x \neq x_0$ with a conical singularity at $x_0$ and deficit angle $2\pi-\theta$; specifically, a sum over the AdS cones $M(2\pi-\theta)$.
 The total action that results is local and the Hayward term describes a local dof on the tip (In Sec. \ref{livingontheedge} we shall revisit this point in more detail), therefore, we can write the asymptotic term in terms of the reparametrization mode which gives a Schwarzian action.  By evaluating this in the dominant solution \eqref{dominant}, and computing the one loop contribution  as usual \cite{Mertens2019}, finally results
 \be
  Z[\phi_\partial, \beta, \theta, \Phi(x_0) ]= \; 
\,\int \left[Dg(x)\right]_{M(2\pi-\theta)} \;\,e^{ \;  \,\frac{(2\pi - \theta(x_0))\,\Phi(x_0) }{8 \pi G_N}}\;  e^{- \,\frac{1}{8 \pi G_N} \int_{\partial M} \sqrt{h} \phi_\partial (K-1)} \ee

\be\nonumber =\left(\frac{\phi_b}{16\pi G_N\beta}\right)^{1/2}\,e^{\;  \;\frac{(2\pi - \theta(x_0))\,\Phi(x_0) }{8 \pi G_N}}\; e^{ -C[\phi_b] \, \frac{\theta^2}{\beta} }\;  .  \ee
Notice the contrast of this expression with the standard formula \eqref{ZJT-valued}.

\subsubsection*{Analysis of (ii)}
 
In agreement with the standard measure of JT gravity \eqref{Dphi}, one can integrate \eqref{ZJT-bdy-defect} over all values of $\Phi(x_0)$, and interpret this as summing over all the fixed area states \cite{Dong2018}, giving place to a new partition function
 \be\label{granZdef} {\cal Z}[\phi_\partial, \beta, \theta]   \equiv \int^{+\infty}_{-\infty}d\Phi(x_0) \, Z[\phi_\partial, \beta, \theta,  \Phi(x_0)],\ee
which can be rewriten as
  \be\label{granZ}
 {\cal Z} [\phi_\partial, \beta, \theta]\equiv 
 \int\left[D\Phi(x)\right]_{M_0}  \left[Dg(x)\right]_{M_0} \; P[\theta]\;  e^{-\frac{1}{16 \pi G_N} \int_{M_0} \; \sqrt{g}\Phi (R+2)\, - \,\frac{1}{8 \pi G_N} \int_{\partial M} \sqrt{h} \phi_\partial (K-1)}, \;
  \ee 
where we have integrated first in $\Phi(x_0)$, and thus the local tip operator is (we take a Wick rotation $\Phi\to i\Phi$)
\be \label{Cconstraint}
P[\theta] \equiv \int^{+\infty}_{-\infty}d\Phi(x_0) \,e^{ \;  \,\frac{(2\pi - \theta(x_0))\,\Phi(x_0) }{8 \pi G_N}}\,= \, \delta\left( 2\pi-\theta(x_0)\,\right),
\ee
which projects the measure $$\left[D\Phi(x)\right]_{M_0}  \left[Dg(x)\right]_{M_0} \;\to \left[D\Phi(x)\right]_{M(2\pi-\theta)}  \left[Dg(x)\right]_{M(2\pi-\theta)} \;$$
This shows that in the resulting gravitational path integral \eqref{granZ} we must sum over \emph{smooth} geometries ($\theta=2\pi$), giving place to a pure gravitational partition function without any conical defect. So in this case, the saddle geometry is the disk.
The partition function can be valued similarly to \eqref{ZJT-valued} but using the constraint \eqref{Cconstraint} to obtain
\be\label{ZJT-Cvalued}
{\cal Z}_{JT}[\phi_\partial, \beta] =\left(\frac{\phi_b}{16\pi G_N\beta}\right)^{3/2}\; e^{ -C[\phi_b]\, \frac{(2\pi )^2}{\beta} }.
\ee 
Note that as expected from the fact that the remaining geometry is a disk we have the partition function obtained in \cite{Saad2019}.

\subsubsection{Random Matrix model and Spectral density}

The physical properties of JT gravity can be captured by a doubly scaling limit of random matrix models \cite{Saad2019}. So, from the analysis above it is now straightforward to compute the spectral density, $\Omega(E)$, for the hypothetical matrix model that is dual to the solution with the Hayward term. It can be read from the expression for the partition function of the theory
\begin{equation}
Z[\beta]=\int_0^\infty dE\, \Omega(E)\,e^{-\beta E},\nonumber
\end{equation}
by taking an inverse Laplace transform. The results reported here holds in the limit when $\Phi_0$ is very large, otherwise higher genus topologies and non-perturbative corrections are important. Because of this we will restore the contribution of the topological term along this discussion.
 
The partition function is a modification of the one obtained for bulk defects in \cite{Mertens2019, Witten2020b}. Then, in our model we get \eqref{ZJTH-fixedA}
\begin{equation}
Z[\beta]=\left(\frac{\phi_b}{16\pi G\beta}\right)^{1/2}e^{\frac{\theta^2}{16\pi G\beta}\phi_b-\frac{(\theta-2\pi)}{8\pi G}(\Phi_0+\Phi_\Gamma)+\frac{\Phi_0}{4G}},\label{ZJTH}
\end{equation}
from which the spectral density can be obtained
\begin{equation}
\Omega_{\Phi_\Gamma}(E,\theta)=e^{\frac{2 \pi  (2 \Phi_0+\Phi_\Gamma)-\theta  ( \Phi_0+\Phi_\Gamma)}{8 \pi  G}}\frac{\sqrt{\phi_b}}{4\pi \sqrt{G E}}\cosh\left(\frac{\theta\sqrt{E \phi_b}}{2\sqrt{G\pi}}\right)\;.\nonumber    
\end{equation}
This spectral density has the same dependence in energy dependence as the one obtained for bulk defects.

Note that if we compute the density of states obtained from the partition function \eqref{granZ} after integration in $\Phi(x_0)$ \eqref{ZJT-Cvalued} we recover the result for the disk \cite{Saad2019}
\begin{equation}
\Omega(E)=\frac{\phi_b}{32 G\pi^{5/2}}\sinh\left({\sqrt{\frac{\pi\phi_b E}{G}}}\right).\nonumber
\end{equation}

\subsection{Classical solutions in JT gravity with/without conical defect}

One of the main conclusions of the analysis above is that JT gravity formulated on Pacman geometries yields the same eoms (and solutions) than the standard analysis
of JT gravity with source term. Nevertheless their \emph{on-shell} actions are subtly different and it has consequences in some computations as we will see in forthcoming sections. In fact as argued before, the eom $R+2=0$ on the interior of $M_P(2\pi-\theta)$ becomes the distributional equation
$$
R+2= 2 (2\pi-\theta) \delta(x-x_0),
$$
as the Pacman closes and the base manifold becomes the conifold $M$ with deficit $2\pi - \theta$.
The most general classical solutions of this system can be written in Schwarschild coordinates as 
\begin{equation}\label{generalsolJT}
ds^2=(r^2-r_0^2)d\tau^2 + \frac{dr^2}{r^2-r_0^2}, \qquad \qquad \Phi(r)= r \,\phi_b, \qquad\qquad r>r_0, \quad \tau\in[0, \beta] \;.
\end{equation}
The metric \eqref{metricAdS2} can be obtained from this one by a re-scaling of the coordinates $r \to  \frac{r}{r_0}$ and $ \tau \to \tau r_0$, and the periods are related by $\theta = \beta r_0$.
We choose this period as coincident with the circumference of the circular solution \eqref{dominant}.

Notice that $\partial_\tau $ is a Killing vector and the horizon is on $r=r_0$. A particular solution is completely determined by three constants/data: $BC \equiv (\phi_b \,, r_0\,, \beta)$. 

Near the horizon, this geometry can be mapped to an ordinary cone with metric \footnote{
By doing a new redefinition of the radial coordinate $\rho^2 =  r^2 - 1 $ we get the solution 
\begin{equation}\label{SystemJTtau}
ds^2=\rho^2 d\tau^2 + \frac{d\rho^2}{\rho^2 + 1} \qquad \qquad \Phi(\rho)= r_0 \, \phi_b \,\sqrt{\rho^2 + 1} \, \qquad \rho >0 \quad \tau\in[0, \theta =r_0\beta] \;.
\end{equation} and for $\rho\approx 0$ ($r \approx 1$) the metric approaches 
\begin{equation}
ds^2=\rho^2 d\tau^2 + d\rho^2 \qquad \qquad \Phi(\rho) \approx \, r_0 \, \phi_b \qquad\qquad \rho >0 \quad \tau\in[0, \theta =r_0\beta] \;.
\end{equation} which proves that there is no conical singularity iff $\theta (\equiv r_0 \beta) = 2\pi$.}
\begin{equation}\label{SystemJTapprox}
ds^2=\rho^2 d\tau^2 + d\rho^2 \qquad  \qquad \Phi(r)= \phi_b\sqrt{\rho^2+1}, \qquad 1\gg\rho > 0, \quad \tau\in[0, r_0 \,\beta] \;,
\end{equation}
implying that the period near the tip/horizon is given by
\be\label{theta}\theta =  r_0 \,\beta .\ee 
This is the reason why $\theta$ is the so-called opening angle, and $\alpha\equiv 2\pi-\theta $ is the \emph{deficit angle}.
Therefore, the cone is a disk without conical singularity iff $\theta=2\pi$.
The presence of a conical defect in JT gravity is intimately related to the possibility of giving arbitrarily the value of the field in two places e.g. on the asymptotic boundary $\phi_b$, and on the horizon $\Phi_\Gamma \equiv \Phi(x_0)$, the point where \emph{there is} a conical singularity.
This is what could be interpreted as an extra dof at the horizon (edge), apart from the boundary mode.

In fact, since $\Phi_\Gamma = r_0 \phi_b $, then the parameter $r_0$ in the solution can be written
\begin{equation}\label{horizonteposition}
 r_0 = \frac{\Phi_\Gamma}{\phi_b},
\end{equation}
and the solution \eqref{generalsolJT} can be completely determined from the data: $BC^\star\equiv (\phi_b \, , \Phi_\Gamma , \beta)$, which are the arguments of the partition function \eqref{ZC-fa-M0}.

So, the opening/deficit angles of the geometry result \emph{dynamically} determined by these data \footnote{In path integral language, ``dynamically'' means that $\langle\alpha\rangle = - \frac{\partial \log Z[BC^\star]}{\partial \Phi_\Gamma} $ gives \eqref{theta-dinamico} as one uses the saddle point approximation.}
\be\label{theta-dinamico} \theta= \frac{\Phi_\Gamma}{\phi_b} \beta, \qquad  \qquad \alpha =2\pi - \frac{\Phi_\Gamma}{\phi_b} \beta .
\ee
This shows that the JT case is consistent with the analysis of the canonical variables in gravity with Hayward term in arbitrary dimensions \cite{Takayanagi2019}, since given $\Phi_\Gamma$(/area) one obtains $\alpha/\theta$ dynamically and vice-versa, because they are canonically conjugate variables.

According to the analysis of the previous section, one can integrate out the number $\Phi_\Gamma$ in the Path integral \eqref{granZ}, and then results the constraint $\alpha=0$ which implies that the saddle manifold is a disk without conical singularity. In that case is very well known that the solution is given just by giving two parameters/data: $(\phi_b\, , \beta)$. In that case $\alpha=0$  $(\theta=2\pi)$, and then using \eqref{theta} and \eqref{horizonteposition}

\begin{equation}\label{MADr0}
 r_0 = 2\pi/\beta \qquad\qquad\Rightarrow \qquad\qquad \Phi_\Gamma = \frac{2\pi}{\beta} \phi_b  \;,
\end{equation} 
and the metric is
\begin{equation}\label{SystemJTsmooth}
ds^2=\left(r^2- \left(\frac{2\pi}{\beta}\right)^2\right)d\tau^2 + \frac{dr^2}{r^2- \left(\frac{2\pi}{\beta}\right)^2}, \qquad \qquad \Phi(r)= r \phi_b, \qquad\qquad r>\frac{2\pi}{\beta}\qquad \tau\in[0, \beta] \;.
\end{equation}
Here it is worth add a note on the relation between the cases discussed in \textbf{(i)}, \textbf{(ii)} and the point of view explained in Sec. \ref{bipartite}. In general \be\label{eomNG}\Phi_\Gamma =  \text{min} \{ \Phi(x) \;\; x\in M\},\ee 
is a condition extra satisfied by the classical solution $\Phi_\Gamma=\Phi(x_0)$, this could be obtained as an extra eom by considering an alternative definition to the gran ensemble \eqref{granZdef} as
\be\label{ZconNG} {\cal Z} \equiv \int_M \sqrt{g}[d^2 x_\Gamma]  Z[BC^\star] = \int_M \sqrt{g}[d^2 x_\Gamma] e^{\Phi_\Gamma \frac{2\pi -\theta}{8\pi G}} \,e^{-I_{JT}[BC^\star] }
\ee
where \emph{a priori}, the \emph{embedding} $x_\Gamma : \Gamma \to M$ is also considered an independent variable in the partition function. Then if the deficit is positive, the leading contribution to  this path integral is giving by \eqref{eomNG}, whose solution is $x_\Gamma =x_0$ (i.e, the minimum of the Dilaton occurs for the position of the tip $x_0$ of the spacetime). 

This formulation coincides with our original proposal \cite{Botta2020} and the arguments of Sec. \ref{bipartite}, and moreover, it is necessary to recover the Dong's recipe to compute R\'enyi entropies.  This can be interpreted as a Nambu Goto theory for the (pointlike) "cosmic brane" on a 2d backreacted spacetime.
Furthermore, the perturbative analysis of \cite{Witten2020b} remarkably leads to a path integral with the same integration measure.  It would be interesting to perform the same analysis with the new Hayward term in future studies.

The reader can verify that this formulation is subtly equivalent to \textbf{(ii)} to recover the (smooth) gravitational path integral, since using the classical relation between $\theta$ and $\Phi_\Gamma$ 
the correct saddle point of \eqref{ZconNG} requires $\theta =2\pi$ as \eqref{Cconstraint}. This point shall be developed in more detail in a  future work.

\section{The replica trick in JT gravity with a Hayward term}\label{replica}

The computation of the R\'enyi entropy $S_n$ and Refined R\'enyi entropy $\hat S _n$ require to compute the partition function associated to $n$ powers of the density matrix $\rho$,
\be\label{Zn-def}
Z_n(\beta) \equiv \text{Tr}\, \rho^n.
\ee 
Then, the analytical extension of this to real values of $n$ around $n=1$ allows to take the limit $n\to 1$ and compute the von Neumann entropy. Moreover, as has been noticed in \cite{Botta2020}, if such analytical extension can be done to purely imaginary values $n\to is$ we obtain the \emph{modular flow}, and one can directly compute its generator: the modular Hamiltonian. This will be the subject of the next section.

We are interested first in reproduce the holographic computation of both spectra $S_n , \hat S _n$ in two set-ups: Fursaev and MLD scenarios in order to understand the replica procedure in 2d gravity. In doing this we will show that both constructions are in conflict with the traditional way of considering defects in JT gravity and these problems can be cured with a Hayward term.

The replica technique consists in consecutively gluing $n$ equal boundaries intervals $[0, \beta]$, and then the trace operation of \eqref{Zn-def} corresponds to identify the extremes of the interval such that the boundary of the geometry is a circle $S^1$ of circumference $n\beta$.
The question is how in different scenarios the opening angle $\theta$ depends on $n$, giving place to different R\'enyi spectrum.
For instance the Fursaev set up \cite{Fursaev2006} is characterized by conifolds whose opening angle spectrum is $\theta(n) = 2\pi n$, such that for $n=1$ the corresponding geometry is the disc ($\theta(1) =2\pi$), while the MLD family is $\theta(n)= 2\pi\;,\; \; \forall n$ (the replicated geometry $M_n$ is smooth).

The von Neumann entropy measures the entanglement of a physical system in a given state and for a specific subset of degrees of freedom. The celebrated Ryu-Takayanagi (RT) \cite{Ryu2006} formula is a powerful tool to compute it in quantum field theory in the context of the gauge/gravity correspondence. This generalizes the Bekenstein-Hawking law for the thermodynamic entropy of Black Holes \cite{Bekenstein1973, Hawking1975} and tell us that the entanglement entropy is given by a quarter of the area of the minimal surface embedded in the dual higher dimensional spacetime with gravity.
Since its discovery  evidence of its validity had been collected (see \cite{Rangamani2016} for a review), and it was finally been derived by computing the gravitational entropy with different replica methods \cite{Fursaev2006, Aitor2013}.

The R\'enyi entropies are a generalization of the von Neumann entropy labeled by an integer $n$,
 \be\label{defirenyi} S_n \equiv \frac{1}{1-n} \log{ \text{Tr} \rho^n },\ee
such that the standard von Neumann entropy $S\equiv-\text{Tr}\rho\log \rho$ is recovered in the limit $n\to 1$. There is an alternative family of measures of entanglement entropy related to the R\'enyi entropies (called refined or modular R\'enyi entropies \cite{Nishioka2018}), given by 
\be\label{defirenyi-hata} \hat{S}_n \equiv -n^2 \partial_n \left(\frac{1}{n} \log{ \text{Tr} \rho^n }\right)= \left(1 \,-\, n \partial_n \right)\,\log{ \text{Tr} \rho^n },\ee
that also coincides with the von Neumann entropy as $n\to 1$. This entropy is specially easy to interpret in terms of thermodynamics \cite{Nakaguchi2016}, in fact, the last equality of \eqref{defirenyi-hata} is the definition of the thermodynamic entropy in the canonical ensemble, but valued on the inverse temperature $\beta=2\pi n$.

A similar area-law prescription for these entropies has been provided \cite{Dong2016}, but in this case the extremal surface interacts with the background spacetime through a tension that depends on the replica index in a specific way
\be
T_n = \frac{n-1}{4n\,G_N}.
\ee 
As we explained in the previous section the formula obtained for the effective action (free energy) has an additional term wrt the traditional computation that comes from the corner in the Pacman geometry
\begin{equation}\label{IJTH}
I[M(2\pi-\theta)] = -\frac{\phi_b\;\theta^2}{16 \pi G \beta}+\frac{2\pi-\theta}{8 \pi G} (\Phi_0 + \Phi_\Gamma).
\end{equation}
Here we will check that this is the right formula to recover known results for the R\'enyi entropies.

\subsection{R\'enyi entropies}

\subsubsection{Fixed area sectors, Fursaev saddle and flat spectrum}

In our approach these three concepts are related in the same scenario that we are going to describe.
We call the Fursaev set up to solutions to BCs: $(\phi_b,\Phi_\Gamma)$,
where the condition on $\Phi_\Gamma $ is nothing but fixing the \emph{area} sector, and the circumference of the  replicated boundary is $\beta_n \equiv n\beta$. So $(\phi_b,\Phi_\Gamma)$ are fixed and independent of this replica number, then using that $r_0 = \Phi_\Gamma / \phi_b$ is independent on $\beta$, then it is also independent on $n$.
Thus the Fursaev 's geometries $F_n$, are the saddle point configuration with these BCs, which are conifolds with opening/deficit angle dynamically determined by these data,
\be \theta= \frac{\Phi_\Gamma}{\phi_b} n \beta,\nonumber
\ee
and metric (with Dilaton)
\begin{equation}
ds^2=\left(r^2- \left(\frac{\Phi_\Gamma}{\phi_b}\right)^2\right)d\tau^2 + \frac{dr^2}{r^2- \left(\frac{\Phi_\Gamma}{\phi_b}\right)^2}, \qquad \qquad \Phi(r)= r \phi_b, \qquad\qquad r>\left(\frac{\Phi_\Gamma}{\phi_b}\right), \quad\qquad \tau\in[0, n \beta],
\end{equation}
where the radial coordinate is bounded by the position of the entanglement surface/defect
\begin{equation}\label{}
 r_0 = \frac{\Phi_\Gamma}{\phi_b} = \,\text{independent on } \, n \;.\nonumber
\end{equation}

Then substituting $\theta_n =r_0 \beta n$, $r_0 \beta \equiv \theta_0$ in the metric and demanding that for $n=1$ there is no conical singularity, we have that $r_0 \beta =\frac{\Phi_\Gamma}{\phi_b} \beta = 2\pi$, therefore we have that (in the Fursaev set up) the opening angle on the tip is $\theta = 2\pi n$.
By taking the quotient of the geometry \eqref{SystemJTsmooth} notice that in this case  the fundamental domain is the solution for $n=1$: $F_1 = F_n/Z_n$.
In other words, the solution \eqref{SystemJTsmooth} are $n$ consecutive copies of the same manifold glued consecutively. 
The replicated solution $F_n\;, n>1$ is \emph{singular} at the tip $\Gamma$, and by evaluation of the on-shell action \eqref{IJTH} reads

\be\label{IFn}
I[F_n]= C \,(2\pi)n \,+\, \frac{2\pi (1-n)}{8\pi G} (\Phi_0+\Phi(x_0)),\qquad\qquad \theta = 2\pi n,\qquad\qquad r_n= 1.
\ee
Therefore, it is easier to compute first the refined R\'enyi entropy 
\be
\hat S _n = \frac{\Phi_0 + \Phi_\Gamma}{4G}.\nonumber
\ee
This is the so called \emph{flat spectrum}. 
The remarkable observation here is that the formula for the action  \eqref{IJTH} in fact  \emph{ requires} the second (Hayward) term to obtain the correct result.

\subsubsection{Smooth solutions,  MLD saddle/spectrum and the Dong prescription}

Once more, here we are going to show many results that only agree with the previous literature \emph{if} we take into account the last Hayward term in the action \eqref{IJTH} which in the standard computations is absent. In particular we shall verify that the Dong's prescription \cite{Dong2016} works correctly in JT gravity\footnote{To our knowledge, this has not been verified in its original form for the JT case. However, other similar formulations that add explict cosmic ``brane'' term have been recently probed in the context of replica wormholes \cite{Almheiri2019}. } with the ingredient of solving the Nambu-Goto action coming from the Hayward tip term. We will begin describing the replica trick in the MLD approach.

We define the MLD  boundary conditions to fixing $\phi_b $ on $[0,\beta]$, repeated $n$ times on the respective copies of the boundary, and the circumference of the  replicated boundary is $\beta(n) \equiv n\beta$. Notice that in this case one \emph{does not fix the area sector} by giving $\Phi_\Gamma$, but it shall be fixed dynamically. As we explained before (Sec. \ref{sources}), these conditions imply that the JT path integral  $Z(\phi_b, \beta)$ is given by \eqref{ZJT-Cvalued}, corresponding to an ensemble where one sums over $\Phi_\Gamma$\footnote{Alternatively, one could fix the deficit angle, that in the ``cosmic brane'' set up is nothing but fixing the tension.}.
Then the saddle point approximation implies \emph{to minimize} with respect to (the JT area) $\Phi_\Gamma$, in agreement with the Ryu-Takayanagi-Dong prescription. This imposes a constraint of smoothness $\alpha=0$ on the saddle geometries $M_n$ expressed by
\be \label{Mnsmooth}
\theta = 2\pi \;\; \forall n \qquad\Leftrightarrow \qquad \beta(n) \, r_0(n) = 2\pi
\ee
Thus, we obtain the positions of the extremal \emph{point} $\Gamma$ and the corresponding MLD 
spectrum \cite{Dong2016, Dong2018, Botta2020}
\begin{equation}\label{}
 r_0(n) = \frac{2\pi}{n \beta} = \frac{\Phi_\Gamma}{\phi_b} \qquad\Leftrightarrow \qquad \frac{ \phi_b\,2\pi}{n\, \beta} = \Phi_\Gamma\nonumber
\end{equation}
and the geometry of $M_n$ is described by the metric
\begin{equation}\label{MnJTsmooth}
ds^2=\left(r^2- \left(\frac{2\pi}{n\, \beta}\right)^2\right)d\tau^2 + \frac{dr^2}{r^2- \left(\frac{2\pi}{n\, \beta}\right)^2},  \qquad \Phi(r)= r \phi_b, \qquad r>\left(\frac{2\pi}{n\, \beta}\right), \qquad \tau\in[0, n \beta] \;.
\end{equation}
The on-shell action is
\be\label{IMn}
I[M_n]=
-\frac{(2\pi)^2 C}{n \beta}.
\ee
According to the definition of replicas and the semi-classical approximation we have 
 \be\label{I-trrhon} I[M_n] \approx \log{ \text{Tr} \rho^n }\ee
 then using \eqref{defirenyi} we obtain a wrong result
 \be
S_n = \frac{1}{1-n} \frac{(2\pi)^2 C}{n \beta}. \nonumber
\ee
The reason is that the direct calculus of $S_n$ involves infinities that, with a suitable regularization cancel out, and one can obtain the correct result. However,  one can achieve this if computes first the refined R\'enyi entropy 

\be\label{defirenyi-hat} \hat{S}_n \equiv -n^2 \partial_n \left(\frac{(2\pi)^2 C}{n^2 \beta}\right)= -\frac{2(2\pi)^2 C}{n \beta}=
\frac{2(2\pi)^2 }{n \beta} \frac{\phi_b}{16\pi G_N} =  \frac{\Phi_\Gamma(n)}{4 G_N} ,
\ee
where in the last equality we have used that the Dilaton at the horizon is: $\Phi_\Gamma (n) = \frac{2\pi \phi_b}{n\beta} $.
From this expression, the R\'enyi entropies can be obtained by integrating the identity
\be\label{renyihat-renyi} \hat{S}_n = n^2 \partial_n \left(\frac{n-1}{n} S_n\right), \ee
and the result is
\be
S_n=\frac{\phi_b}{4G}\frac{\pi}{\beta}\frac{n+1}{n}.\label{renyi}
\ee
Nevertheless we have not even used the Dong prescription. This is an interesting point for analysis. 

By taking the quotient of the geometry \eqref{MnJTsmooth} we obtain the fundamental domain $\hat M_n = M_n/Z_n$ whose boundary is $S^1_\beta$, and 
the metric (+ Dilaton) of this space is the same as \eqref{MnJTsmooth} but on the range of coordinates

\begin{equation}\label{hatMnJTsmooth-range}
 r >\left(\frac{2\pi}{n\, \beta}\right), \qquad\qquad\qquad \tau\in[0,  \beta],
\end{equation}
while the opening angle now is $\hat \theta = r_0(n)\, \beta = \frac{2\pi}{n\, \beta} \, \beta = \frac{2\pi}{n}$, and according to our general expression \eqref{IJTH}   the  on-shell action is
\be\label{hatM}
I[\hat M _n]= C \frac{(2\pi)^2}{n^2 \beta} \,+\, \frac{2\pi}{8\pi G}\frac{n-1}{n} (\Phi_0+\Phi(x_0)) ~,~~~~~~\;  \;\;\; r_0(n) =\frac{2\pi}{n\, \beta}.
\ee
Then, the solution \eqref{MnJTsmooth} is given by $n$ copies of $\hat M_n$ glued consecutively. Comparing with \eqref{IMn}, the first term of traditional JT gravity is additive and satisfies $$I_{JT}[M_n]= n I_{JT}[\hat{M}_n].$$
Equation \eqref{hatM} is the JT action of gravity plus (coupled to) a  Nambu-Goto term describing a cosmic brane with tension $T = \frac{n-1}{4Gn}$ as the model \cite{Dong2016}. Remarkably we have derived it from a Hayward term present in the original Pacman geometry.

As an aside, note that the capacity of entanglement (a quantum information quantity analogous to the heat capacity in thermodynamics) can be written as 
\begin{equation}
C=\partial_n^2(\log Tr\rho^n)|_{n=1}=\langle K \rangle_n^2-\langle K^2 \rangle_n,\nonumber
\end{equation}
where $K$ is the modular Hamiltonian and the expectation value is taken w.r.t. the state $\rho_n\equiv\rho^n$. So, the capacity of entanglement measures quantum fluctuations of the modular Hamiltonian. In \cite{Verlinde2019} was claimed (for the vacuum state) that $\langle K^2 \rangle-\langle K \rangle^2=\frac{A(\Sigma)}{4G_N}$ ($\Sigma$ is a Rindler horizon) for any strongly coupled $CFT$ with a large $N$ gravitational dual that is described by the Einstein action. Here note that using \eqref{renyi} the capacity gives $C=\frac{\phi_\partial}{2G}\frac{\pi}{\beta}$, which using \eqref{MADr0} gives $C=\frac{\phi_\Gamma}{4G}$ and is in agreement with the expectation of \cite{Verlinde2019}.

\section{Replica symmetry, the modular flow and area operator}\label{modular}

In this section we will evaluate the modular flow (and the modular Hamiltonian) from the computation of the $n-$th power of the density matrix, $\rho^n$. We shall use a path integral expression for the matrix elements of the \emph{bulk representations} of this state, that obviously involves Hayward term associated to the $n$-dependent opening angle $\theta(n)$, and as a result, we'll obtain the \emph{area} operator \cite{Botta2014, Jafferis2014} as part of the modular Hamiltonian, in agreement with the JLMS conjecture \cite{JLMS}.

Recalling our prescription \eqref{rhoA-osum} for the left (reduced) density matrix, the $n$ power of it preserve the SS sectors structure
\be 
\rho^n(L) = \bigoplus_{\Gamma}\; \rho (n, \Gamma) \;\sim \bigoplus_{\Gamma}\;d(\Gamma)\; e^{-\,n \,K(\Gamma)}\;. \ee
The left hand side of this expression stands for the boundary quantum theory, while the right hand side refers to the density matrix in (JT) gravity.
Since each sector is labeled by the position of the entangling point $x_\Gamma$, and since the Dilaton is considered an observable (see Sec. \ref{bipartite}), they can also be labeled by $\Phi_\Gamma$ and referred to as fixed area sectors.
So these density matrix elements are functionals: $\rho[\Phi_\Gamma, \phi^\pm]= \langle \phi^+ | \rho (\Gamma) |\phi^-\rangle$. In the mindset of Sec. \ref{sources}, the suitable path integral expression for each \emph{block} of \eqref{rhoA-n-osum} is
\be\label{rho-fa-M0}
 \langle \phi^+ | \rho (n, \Gamma) |\phi^-\rangle =
 \int_{(\Phi(B^\pm), g(B^\pm)) = \phi^\pm}\left[D\Phi(x)\right]_{M_P} \, \left[Dg(x)\right]_{M_P} \; e^{ \;  \,\frac{(2\pi - \theta(n))\,\Phi_\Gamma }{8 \pi G_N}}\;  e^{\frac{1}{16 \pi G_N} \int_{M_P} \; \sqrt{g}\Phi (R+2)\, + \,\frac{1}{8 \pi G_N} \int_{\partial M_P} \sqrt{h} \phi_\partial (K-1)},
  \ee
  where $\phi^\pm$ denotes the Dirichlet boundary conditions on geodesics $B^\pm$. This is the reason why the GHY terms associated to $B^\pm$ vanish.
For simplicity we assume that $\Phi$  is at least  ${\cal{C}}_0$ (continuous) on the total boundary $\partial M_P$, then $\Phi_\Gamma \equiv \lim_{x\to x_\Gamma} \Phi^\pm(x) \equiv \Phi_\Gamma \;, \;\, x\in B^\pm$. 
 As argued in Sec. \ref{replica}, the saddle geometry is $M_P(1)  = M^- \cup M^+$  smoothly glued on the shared surface ${\bar B}$, whose boundaries are the branches (curves) $B^-$ and $B^+$ with an opening angle $\theta(1)$. As we consider $n$ (consecutively) replicated boundaries, the euclidean geometry is a new Pacman $M_P(n)$
 characterized by a corner with an opening angle $\theta(n)$.

Now, integrating out the Dilaton in \eqref{rho-fa-M0}, results
  
  \be\label{rho-fa-M0-2}
  \langle \phi^+ | \rho (n,\Gamma) |\phi^-\rangle  =
 \int_{(\Phi(B^\pm), g(B^\pm)) = \phi^\pm} \, \left[Dg(x)\right]_{M_P(n)} \; e^{ \;  \,\frac{(2\pi - \theta(n))\,\Phi(x_0) }{8 \pi G_N}}\;  e^{ \,\frac{1}{8 \pi G_N} \int_{\partial M^0_P(n)} \sqrt{h} \phi_\partial (K-1)},
  \ee
  where $M^0_P(n)$ is locally AdS$_2$.
  For geodesic boundaries $B^\pm$: $K=0$, one can define $h^\pm(r) \equiv h(r,\pm \tau)$, whose proper length is
  \be ds_\pm^2 \equiv h^\pm(r)  dr^2 \;,\qquad\qquad 0\leq \tau \leq \theta \equiv \tau^+ -\tau^- \;.\nonumber
  \ee
With a suitable redefinition of the timelike coordinate $\tau$ the metric in the Pacman \emph{saddle} geometry $M^0_P$ is
  \begin{equation}\label{Pacmanmetric}
ds_P^2=(r^2-1)d\tau^2+\frac{dr^2}{r^2-1},\qquad\qquad 1 \leq r < \infty\qquad\qquad 0 < \tau <  \theta = r_0\,\beta\,.
\end{equation}
Let us highlight  that this is not periodic and does not have any conical singularity. Just upon taking trace of the state the  to compute the partition function, the Pacman mouth closes by gluing $B^\pm$, and the geometry turns out to be a (periodic) cone: $M(2\pi-\theta(n))$.

 Therefore, in the same way that has been argued in the Sec. \ref{replica} in the case that the  splitting point of the geometry $\Gamma$ (i.e, the SS sector) is fixed, we must have $r(\Gamma) =r_0= \text{fixed}$, then $\theta = r_0 \beta n = 2\pi n $, and we finally have 
  \begin{align*}
    \langle \phi^+ | \rho (n, \Gamma) |\phi^-\rangle &=
 \int_{(\Phi(B^\pm), g(B^\pm)) = \phi^\pm} \, \left[Dg(x)\right]_{M^0_P} \; e^{ \;  \,\frac{(2\pi - 2\pi n)\,\Phi(x_0) }{8 \pi G_N}}\;  e^{  \,\frac{n}{8 \pi G_N} \int_{\partial M_P(1)} \sqrt{h} \phi_\partial (K-1)} \\  
 &= \;\langle \phi^+ |\,e^{\frac{(2\pi - 2\pi n)\Phi(x_0) }{8 \pi G_N}}\; \rho_{JT}^n (\Gamma) |\phi^-\rangle,
  \end{align*} 
where we have used that 
$$\frac{1}{8 \pi G_N} \int_{\partial M_P(n)} \sqrt{h} \phi_\partial (K-1)= \frac{n}{8 \pi G_N} \int_{\partial M_P(1)} \sqrt{h} \phi_\partial (K-1)\;. $$
Finally, following \cite{Botta2020} the formula for the generator of the modular flow $U(s)$ is
  \be\label{modularKqft}
{\cal{K}} = -\, \lim_{n\to 0}\, U(-i2\pi n)\frac{\partial }{\partial n} U(i2\pi n) \, =  -\, \lim_{n\to 0}\, \rho(-n)\frac{\partial \rho(n)}{\partial n} ,
\ee
for each area fixed sector $\Gamma$. Clearly, because of the factor $\;e^{\frac{(2\pi - 2\pi n)\Phi_\Gamma }{8 \pi G_N}}$ in \eqref{rho-fa-M0-2} the dominant sector (for arbitrary number of replicas $n > 1$) is for the minimum of the function $\Phi_\Gamma$ in the dominant AdS geometry.
Upon quantization, the Dilaton field has to be promoted to an operator, and the result of this calculus is the modular Hamiltonian associated to the (say) left QM system
\be\label{modularKqftJT}
{\cal{K}}(L) = \frac{\hat \Phi_\Gamma }{4 G_N} + 2\pi H_{JT} (L),
\ee
 where $H_{JT}$ is nothing but the (left-side) boundary mode Hamiltonian computed in the literature on JT gravity \cite{Jafferis2019, Yang2018}, and we shall evaluate the matrix elements using the Schwarzian theory. The important conclusion of this part is that, since $\hat \Phi_\Gamma$ is the area operator on JT gravity, this shows the JLMS proposal in the JT context.
  Roughly speaking, the interpretation of the formula \eqref{modularKqft} is that the replica symmetry is nothing but a restriction (to integer numbers) of the continuous symmetry generated by the modular flow $U(s) \equiv \rho^{is}$\,\cite{Botta2020}.

 The derivation above is a computation at level of \emph{operators} and one could test it by computing some related quantity directly. 
 For instance, the expectation value of the modular Hamiltonian under replicas, and/or standard inequalities of information theories. Taking $\beta \equiv 2\pi$ we have 
\be\label{}
\log Z_n \equiv \log \, Tr\, \rho^n = I[M_n]= C \left. \frac{(2\pi)^2 n}{\beta} \right|_{\beta= 2\pi} \,+\, \frac{1-n}{4 G} \Phi_\Gamma\; .\label{eqperC}
\ee
Then, by taking derivative with respect to $n$
\be\label{}
\frac{\partial}{\partial n}  \log Z_n = \frac{1}{Z_n} \, \frac{\partial}{\partial n}  Z_n =
\frac{1}{Z_n} \,Tr\,  \frac{\partial}{\partial n} \, \rho^n =\frac{1}{Z_n} \,Tr\,  \log \,\rho \, \rho^n = -\frac{1}{Z_n} \,Tr\,  {\cal{K}}\,  \rho^n \equiv -\langle {\cal{K}} \rangle_n, \nonumber
\ee
the result is
$$ \langle {\cal{K}} \rangle_n \,=\,-\, (2\pi)C   \,+ \, \frac{\Phi_\Gamma}{4 G} ~=  2\pi\,E_L + \frac{\Phi_\Gamma}{4 G} \;.$$
Which shows that the operator ${\cal{K}}$ contains the areas operator as a term, in agreement with the result \eqref{modularKqftJT}. $E_L$ is the energy of the boundary mode computed from the Schwarzian action.

\section{A note on the asymptotic and edge modes}\label{livingontheedge}

In arbitrary spacetime dimension the Pacman opening angle is associated to the canonical conjugate of the area element of the codim-2 surface $\Gamma$, and this dof at the corner plausibly encodes edge modes in JT gravity \cite{Takayanagi2019}. We have seen that JT gravity can be effectively formulated as a quantum mechanical system with a single asymptotic dof associated to the reparameterization of the time field $\tau(u)$. Here we elaborate a bit more on the edge mode in this Schwartzian theory.

Recall that the time coordinate is nothing but a \emph{local} (scalar) field defined on all the spacetime. Therefore $\tau(u)$ refers to this field \emph{on the points near the asymptotic boundary} ${\cal{C}} \equiv \partial M$, and we call $\tau_\Gamma(u)$ to the same field on $\Gamma$. The corresponding boundary condition for a conifold $M(2\pi - \theta)$ is 
\be \label{BCtau}  \tau(u+\beta)= \tau(u) + \frac{\theta}{r_0},\ee 
and on the corner
\be \label{BCtauGamma}  \tau_\Gamma(u+\beta)= \tau_\Gamma(u) + \frac{\theta}{r_0}.\ee
Thus the Hayward term can be written in this field as
\begin{equation}
I_H (\tau_\Gamma) :=2\Phi_\Gamma (2\pi -\theta_\Gamma ) = 2\int_0^{\beta} du \;\phi_\Gamma  \;\left(\frac{2\pi}{\beta} - r_0 \dot\tau_\Gamma \right). 
\end{equation}
Notice that here $\phi_\Gamma$ plays a similar role than $\phi_\partial$ in the asymptotic term, and it is thought as an independent Dirichlet data. So the total action becomes
\begin{equation}
I_{JTH} (\tau, \tau_\Gamma) = 2\int_\partial \sqrt{h}\Phi(K-1) + 2\Phi_\Gamma (2\pi -\theta_\Gamma ) = 2\int_0^{\beta} du \;\left(\phi_b\;  \text{Sch}\left(\tan\left[\frac{ r_0 \tau }{2}\right];u\right) \; + \;\phi_\Gamma  \;\left(\frac{2\pi}{\beta}-r_0 \dot\tau_\Gamma \right) \right).
\end{equation}
Note that it is written in terms of a Lagrangian in the fields $\tau(u), \tau_\Gamma(u)$.

A surface $B(u)$, $0\leq u\leq \beta$, of the Pacman intersects these two curves, and thus the canonically conjugate momentum associated to the $\tau(r, u)$ field in these two points is
\begin{equation}
\label{pi-asintotico} 
\pi_\tau (\partial M) = \frac{\partial {\cal L}_{JT}({\dddot \tau, \ddot \tau, \dot \tau, \tau )}}{\partial \dot \tau},
\end{equation}
and  on the tip, the momentum remarkably is the Dilaton field
\be\label{pi-tau2} \pi_\tau (\Gamma) =  \Phi_\Gamma,
\ee
as we expected from canonical analysis. We want to stress that eq. \eqref{pi-asintotico} is schematic and only expresses the first momentum field, one should define more fields associated to derivatives of $\tau$ in order to obtain a Lagrangian depending only on these fields and their first time derivative. 

Notice that instead of having $[\theta_\Gamma, \phi_\Gamma]=i/2$, see App. \ref{ADM}, we have the consistent CCR
\be [\tau_\Gamma , \phi_\Gamma] = i/2, \ee
where $\int_0^\beta\, du \,\dot \tau_\Gamma = \tau_\Gamma(\beta) - \tau_\Gamma(0) = \theta/r_0$. The main novelty of this description is that the total (JTH) path integral  \eqref{ZJTH-fixedA} reduces to an integral only over the fields $\tau(u), \tau_\Gamma(u)$, satisfying \eqref{BCtau}-\eqref{BCtauGamma}.
The path integral can then be written as  
\begin{equation}\label{ZJTHtau2}
Z_{JTH} (\phi_b \,, \Phi_\Gamma) = \int [D \tau(u)][D \tau_\Gamma(u)] \exp^{-I_{JTH}(\tau, \tau_\Gamma)},
\end{equation} and $\tau_\Gamma$ is what we can identify wth the JT edge mode, in line with the observations of \cite{Takayanagi2019}. 
In fact notice that \emph{any } arbitrary $\tau_\Gamma(u)$ minimizes the action, which realizes it as parameter for the gauge symmetry associated to the edge.
This implies that $Z= Z(\partial M) Z(\Gamma)$ where $Z(\Gamma)$ becomes proportional to the volume of the space of functions $\tau_\Gamma(u)$, and one recover the results from Sec. \ref{sources}.

\section{Concluding Remarks}\label{conclu}

In the context of the holographic duality, the quantum state of a subsystem is described by the reduced density matrix built up from the Hartle-Hawking wave function by tracing out the complementary dofs. In the semiclassical approximation, the matrix elements of the density operator can be evaluated from the (dominant) solutions of gravity for suitable Dirichlet boundary conditions on Pacman-shaped manifolds. 

In this work we consider these Pacman geometries in JT gravity. As this kind of geometries have a corner in its boundary we added a Hayward term to the action in order to have a well-posed Dirichlet problem. We did the explicit computation in JT gravity and showed the resulting action as well as it on-shell solution.

We study more in depth the hypothesis of \cite{Botta2020} on the von Neumann decomposition of the density operator (and Hilbert space) in block diagonal form, summing over all the possible splittings of bulk dofs as the SS sectors. Since in JT quantum gravity the area ($\Phi_\Gamma$) should be a natural observable of codim 2 surfaces, these are also fixed area sectors.

Indeed, the Hayward term contribution is shown to be relevant to study details on that decomposition, and the edge modes contribution to the entanglement entropy given by the RT formula \cite{Ryu2006}. In particular we gave some constraints on the underlying symmetry group of JT gravity as viewed as a diffeomorphism gauge theory, recover some results obtained previously with different methods \cite{Lin2018}, and obtain some interesting generalizations.

By studying the partition function obtained for closing the Pacman geometry we arrived to a new description for the case with a conical defect, different in spirit to the previously obtained in \cite{Mertens2019, Witten2020a}. We showed that after summing over all the possible values for $\Phi_\Gamma$ the defect contribution disappears and we end up with the standard JT gravity partition function for a disk geometry. Also, we computed the spectral density of states corresponding to this solution and that hypothetically can be the same density of states that corresponds to a dual random matrix model. 

We then compute the R\'enyi and refined R\'enyi entropies in this context, showing results consistent with the literature and pointing out several subtleties in the computation. Using this we obtained from first principles the JLMS relation for the modular Hamiltonian. 

We conclude our work with two possible future directions.
The first would be that of considering 
multiple corners $\Gamma_i$ as boundary condition to the HH state studied in Sec. \ref{HayenJT}. The resulting spacelike surface $\Sigma$ will now contain pieces that do not immediately correspond to either boundary subsystems which would be interesting to study on their own. Moreover, in building a density matrix associated to it, one can consider several intermediate objects upon taking partial traces on different bulk subregions. For two corners $\Gamma_1$ and $\Gamma_2$, one can envision a thermo-mixed double state in the fashion of \cite{Verlinde2020}. In a more general scenario 
and especially in replica trick computations, multiple gluing possibilities arise that spontaneously break the replica symmetry and their possible dominance at different time regimes 
are goals for future research.

We also leave for future work the interesting situation on which we prepare the a state using the Pacman geometry but on the deformed JT gravity theory \cite{Mertens2019, Witten2020a}. This is interesting because in such a case we will end up with geometries with both kinds of defects, bulk and boundary, and the resulting partition function will be very different from those obtained before.

\appendix

\section*{Acknowledgments}
We thanks Horacio Casini and Felipe Rosso for helpful discussions. The authors are supported by CONICET.

\section{Schwarzian Action}\label{schwarz}

Here we will show how to get the Schwarzian action used in Sec. \ref{onsolution}. We start by writing the metric
\begin{equation}
    ds^2=(r^2-r_0^2)d\tau^2+\frac{dr^2}{r^2-r_0^2},\qquad\qquad 0<\tau<\theta/r_0\label{metape}
\end{equation}
which has the asymptotic boundary at $r\rightarrow\infty$. But, we want to cut the spacetime before that and we will parametrize a general boundary curve by the proper boundary time $u$. Using that the curve will be $\gamma(u)=(\tau(u),r(u))$ and the line element over it reads
\begin{equation}
    ds^2|_\partial=\frac{du^2}{\epsilon^2}.\nonumber
\end{equation}
As was explained in the main text the parameter $\epsilon$ is a regulator such that $\epsilon\rightarrow0$. The length of $\gamma(u)$ is $L=\int_0^\beta ds=\frac{\beta}{\epsilon}$ but along the text and as is usual in the literature we will call by $\beta$ to the length of the curve. The tangent vector to $\gamma(u)$ can be writen as 
\begin{equation}
    \Vec{\eta}=r'(u)dr+\tau'(u)d\tau,\nonumber
\end{equation}
and the unit normal vector is
\begin{equation}
    v^\tau=-\frac{r'}{\sqrt{r^2-r_0^2}\sqrt{\tau'^2(r^2-r_0^2)+r'^2}},\,\,\,\,\, v^r=\frac{(r^2-r_0^2)^{3/2}\tau'}{\sqrt{\tau'^2(r^2-r_0^2)+r'^2}}.\nonumber
\end{equation}
Using that the trace of the extrinsic curvature $K=\nabla_\mu v^\mu$ and differentiating the line element we obtain
\begin{equation}
    K=\frac{r-\epsilon r''}{\tau'(r^2-r_0^2)\epsilon},\nonumber
\end{equation}
which can be expanded to second order in $\epsilon$ to get
\begin{equation}
    K=1+\epsilon^2\left(\frac{(2r_0^2-1)\tau'^4-3\tau^4+2\tau'\tau'''}{2\tau'^2}\right)+\ldots=1+\epsilon^2\left(Sch\left[\tan\left(\frac{\sqrt{2r_0-1}}{2}\tau(u),u\right)\right]\right)+\ldots.\label{Kexp}
\end{equation}
The dots represents order $\epsilon^4$ corrections and higher. So,
\begin{equation}
 -\frac{1}{8 \pi G_N} \int_{\partial M} \sqrt{h} \Phi (K-1)=-\frac{\phi_b}{8\pi G_N}\int_0^\beta du\,Sch\left[\tan\left(\frac{\sqrt{2r_0-1}}{2}\tau(u),u\right)\right],\nonumber
\end{equation}
where is clear that the term $\frac{1}{8 \pi G_N} \int_{\partial M} \sqrt{h} \Phi$ was added to the usual Gibbons-Hawking term because of holographic renormalization to cancel the divergent order $\epsilon^{-2}$ term coming from K in \eqref{Kexp}. This is the first term written in \eqref{JTH2} where we used $r_0=1$.
In these coordinates the equation of motion 
\begin{equation}
    \frac{Sch[\tau,u]'}{\tau'}-\tau''=0,\nonumber
\end{equation}
is satisfied by the solution \eqref{dominant}
\be
\tau(u)=\frac{\theta}{\beta}u,\nonumber
\ee
where the constant was fixed to ensure that $\tau(u+\beta)=\tau(u)+\theta$.

\section{ADM analysis and edge modes}\label{ADM}

In order to build the ADM analysis we borrow from \cite{Louis1993, Yang2018}. We stress that the ADM formalism does require a real-time set-up. We rewrite the metric as
\begin{equation}
    ds^2=-N^2 dt^2 + \sigma^2 (dx+N_x dt)^2
\end{equation}
and rewrite the dynamical piece of the action in terms of these functions
\begin{equation}
    S =\int \sqrt{g}\;\Phi\; (R+2) =\int \sqrt{g}\;\Phi \; R + 2\int \sqrt{g}\;\Phi = 2\int \sqrt{g}\;\Phi\; \frac{R^{1}\,_{212}}{g_{22}} + 2\int \sqrt{g}\;\Phi
\end{equation}
where the key relation is that in 2d gravity $R=\frac{2}{g_{22}}R^{1}\,_{212}$. Disregarding boundary terms, one can show that all terms with no derivatives in $\Phi$ other than $2\int \sqrt{g}\Phi$ cancel between each other, leading to,
\begin{equation}
    S = 2\int \frac{\dot \Phi}{M}(-\dot\sigma+(N_x \sigma)') + 2\int \Phi'\left( \frac{M'}{\sigma} + \frac{M_x}{M}(\dot\sigma-(N_x \sigma)') \right)+ 2\int \sqrt{g}\Phi
\end{equation}
where we used that $\sqrt{g}= M \sigma$. One then gets $\Pi_M=\Pi_{N_x}=0$
\begin{equation}
    \Pi_\Phi=\frac{1}{N}(-\dot\sigma+(N_x \sigma)') = K \;\sigma  
    \qquad\qquad 
    \Pi_\sigma=\frac{1}{N}(-\dot\Phi + N_x \Phi') = \partial_n \Phi
\end{equation}
where the timelike normal vector is $n = N^{-1} \{-1 ,N_x\}$, $ n^2 = -1$ and $K=\nabla_\mu n^\mu$ is the asymptotic extrinsic curvature. 

One can then show that the canonical Hamiltonian, again disregarding boundary terms, 
\begin{equation}
    H_{ADM}=\int dx \left(  \Pi_\Phi \dot\Phi + \Pi_\sigma \dot\sigma - {\cal L} \right)=\int dx N {\cal H} + N_x {\cal H}_x
\end{equation}
\begin{equation}
    {\cal H}=-\Pi_\Phi \Pi_\sigma + \frac{\Phi''}{\sigma}- \frac{\sigma \Phi'}{\sigma^2}-\sigma \Phi \qquad\qquad {\cal H}_x=\Pi_\Phi \Phi'-\sigma \Pi_\sigma'\;.
\end{equation}
By combining these constraints, one can define
\begin{equation}
    -\frac{2}{\sigma}(\Phi' {\cal H}+\Pi_\sigma {\cal H}_x)= \left( \Pi_\sigma^2+\Phi^2-\frac{\Phi'\,^2}{\sigma'\,^2}\right)'\equiv S^2 \approx 0
\end{equation}

By choosing the gauge in which the normal derivative of $\Phi$ vanishes, $\partial_n \Phi=\Pi_\sigma=0$, one can explicitly solve the constraint as
\begin{equation}
    \left( \Phi^2-\frac{\Phi'\,^2}{\sigma'\,^2}\right)' \equiv S^2 
    \qquad\Rightarrow\qquad
    \Phi(\sigma x)=S \cosh(\sigma x)
\end{equation}
which explicitly coincides with the classical solution on $\Sigma$. This indicates that the complete information on $\Sigma$ is completely determined by a single number. In particular, depending on the variational problem of interest, this can be either $S$ or $\theta$. On this gauge, the $\theta_b$ angles between $\Sigma$ and the asymptotic boundaries are always $\theta_b=\pi/2$.

A complementary analysis comes from the variation of the dynamical piece of the Lorentzian action. Consider a real time evolution from $t=-\infty$ up to a surface $\Sigma$ containing a cusp $\Gamma$ at a certain point in the interior. As shown above, one can put $\theta_b=\pi/2$ via a gauge fixing, so we can disregard the contributions coming from the asymptotic boundaries. The variation of the Lorentzian action can be written as

\begin{align}
\delta I_D +\delta\left( 2\int_{\Sigma}\!\! \sqrt{h} \Phi K\right)=
  (\text{eoms})+\int_{-\infty}^{\Sigma} dt \frac{ d}{dt}\Theta
  =(\text{eoms})+\int_{\Sigma}\!\! \Pi_\Phi \;\delta\Phi +\int_{\Sigma}\!\! \Pi^{\nu\rho}_h\; \delta h_{\nu\rho} + 
 2 \;  \Phi_\Gamma \delta\theta_\Gamma
\end{align}
which can obtained in a similar fashion as \eqref{eqforAppB} starting from the Lorentzian signature action. In the expression above we have introduced $(\text{eoms})$ to collectively denote the equations of motion, and the symplectic potential $\Theta$ from where one can directly read the dofs of the phase space \cite{Takayanagi2019}. 
The first two terms in $\Theta$ contain the standard codim-1 gravitational dofs one would expect from a Dilaton+gravity problem.
The appearance of a codim-2 term in this expression shows that the $\Gamma$ corner has induced a new set of dof $\theta_\Gamma$ and its conjugate momentum $\Phi_\Gamma$ in the problem. Its commutations relations are ($16\pi G=1$)
\begin{equation}
    [\theta_\Gamma,\Phi_\Gamma]=i/2
\end{equation}
much like the $x,p$ commutation relations in standard QM, disregarding the 2 which can be modified via a parameter redefinition. As stated in the main text, these new dofs arise commonly in gauge theories when splitting the spacetime into subregions. The intuition is that there are infinite non-vanishing gauge transformations that mix the interior and exterior that impede a straightforward splitting. One can see these new codim-2 dofs as a manifestation of the broken gauge symmetries becoming physical dofs in each subsystem \cite{Takayanagi2019}. 

Two comments are due. The first is that technically one can also include the topological piece of the action in our analysis and define the canonical momentum as $\Phi_0+\Phi_\Gamma$. Since $\Phi_0$ is a constant for all our present purposes, this just represents a rigid displacement on the conjugated momentum, which is unimportant. As a final comment, notice that $\theta_\Gamma$ in this Lorentzian context is a boost rather than a proper angle as in the Euclidean approach in most of the main text. Their connection is straightforwardly made via a Wick rotation $\theta_\Gamma\to i \theta_\Gamma$.

\pagebreak


\begin{thebibliography}{99}
\bibitem{Botta2020}
M.~Botta-Cantcheff, P.~J.~Martinez and J.~F.~Zarate,
``R\'enyi entropies and area operator from gravity with Hayward term,''
JHEP \textbf{07} (2020) no.07, 227
doi:10.1007/JHEP07(2020)227
[arXiv:2005.11338 [hep-th]].

\bibitem{Jafferis2019}
D.~L.~Jafferis and D.~K.~Kolchmeyer,
``Entanglement Entropy in Jackiw-Teitelboim Gravity''
 [ arXiv:1911.10663 [hep-th] ].

\bibitem{Jackiw1984}
R.~Jackiw,
``Lower Dimensional Gravity,''
Nucl. Phys. B \textbf{252} (1985), 343-356\\
doi:10.1016/0550-3213(85)90448-1

\bibitem{Teitelboim1983}
C.~Teitelboim,
``Gravitation and Hamiltonian Structure in Two Space-Time Dimensions,''
Phys. Lett. B \textbf{126} (1983), 41-45
doi:10.1016/0370-2693(83)90012-6

\bibitem{Penington2019}
G.~Penington,
``Entanglement Wedge Reconstruction and the Information Paradox,''
JHEP \textbf{09} (2020), 002
doi:10.1007/JHEP09(2020)002
[arXiv:1905.08255 [hep-th]].

\bibitem{Marolf2019}
A.~Almheiri, N.~Engelhardt, D.~Marolf and H.~Maxfield,
``The entropy of bulk quantum fields and the entanglement wedge of an evaporating black hole,''
JHEP \textbf{12} (2019), 063
doi:10.1007/JHEP12(2019)063
[arXiv:1905.08762 [hep-th]].



\bibitem{Mahajan2019}
A.~Almheiri, R.~Mahajan and J.~Maldacena,
``Islands outside the horizon,''
[arXiv:1910.11077 [hep-th]].

\bibitem{Prem2020}
T.~J.~Hollowood and S.~P.~Kumar,
``Islands and Page Curves for Evaporating Black Holes in JT Gravity,''
JHEP \textbf{08} (2020), 094
doi:10.1007/JHEP08(2020)094
[arXiv:2004.14944 [hep-th]].



\bibitem{Lin2018}
J.~Lin,
``Entanglement entropy in Jackiw-Teitelboim Gravity,''
[arXiv:1807.06575 [hep-th]].

\bibitem{Lin2021}
J.~Lin,
``Entanglement entropy in Jackiw-Teitelboim gravity with matter,''
[arXiv:2107.11872 [hep-th]].

\bibitem{Saad2019}
P.~Saad, S.~H.~Shenker and D.~Stanford,
``JT gravity as a matrix integral,''
[arXiv:1903.11115 [hep-th]].

\bibitem{Kitaev2015}
A.~Kitaev, 
``A simple model of quantum holography,''
Seminar at KITP, 2015.

\bibitem{Polchinski2016}
J.~Polchinski and V.~Rosenhaus,
``The Spectrum in the Sachdev-Ye-Kitaev Model,''
JHEP \textbf{04} (2016), 001
doi:10.1007/JHEP04(2016)001
[arXiv:1601.06768 [hep-th]].



\bibitem{Maldacena2016}
J.~Maldacena and D.~Stanford,
``Remarks on the Sachdev-Ye-Kitaev model,''
Phys. Rev. D \textbf{94} (2016) no.10, 106002
doi:10.1103/PhysRevD.94.106002
[arXiv:1604.07818 [hep-th]].

\bibitem{Hayward1993}
  G.~Hayward,
  ``Gravitational action for space-times with nonsmooth boundaries,''
  Phys.\ Rev.\ D {\bf 47} (1993) 3275.
  doi:10.1103/PhysRevD.47.3275

\bibitem{Takayanagi2019}
T.~Takayanagi and K.~Tamaoka,
``Gravity Edges Modes and Hayward Term,''
JHEP \textbf{02} (2020), 167
doi:10.1007/JHEP02(2020)167
[arXiv:1912.01636 [hep-th]].

\bibitem{Dong2016}
X.~Dong,
``The Gravity Dual of R\'enyi Entropy,''
Nature Commun. \textbf{7} (2016)
[arXiv:1601.06788 [hep-th]].

\bibitem{Dong2018}
X.~Dong, D.~Harlow and D.~Marolf,
``Flat entanglement spectra in fixed-area states of quantum gravity,''
JHEP \textbf{10} (2019), 240
doi:10.1007/JHEP10(2019)240
[arXiv:1811.05382 [hep-th]].



\bibitem{Almheiri2019}
A.~Almheiri, T.~Hartman, J.~Maldacena, E.~Shaghoulian and A.~Tajdini,
``Replica Wormholes and the Entropy of Hawking Radiation,''
JHEP \textbf{05} (2020), 013
doi:10.1007/JHEP05(2020)013
[arXiv:1911.12333 [hep-th]].

\bibitem{Ellerin2021}
S.~Colin-Ellerin, X.~Dong, D.~Marolf, M.~Rangamani and Z.~Wang,
``Real-time gravitational replicas: Low dimensional examples,''
[arXiv:2105.07002 [hep-th]].

\bibitem{Ryu2006}
S.~Ryu and T.~Takayanagi,
``Holographic derivation of entanglement entropy from AdS/CFT,''
Phys. Rev. Lett. \textbf{96} (2006), 181602
doi:10.1103/PhysRevLett.96.181602
[arXiv:hep-th/0603001 [hep-th]].

\bibitem{Bekenstein1973}
J.~D.~Bekenstein,
``Black holes and entropy,''
Phys. Rev. D \textbf{7} (1973), 2333-2346
doi:10.1103/ PhysRevD.7.2333

\bibitem{Hawking1975}
S.~W.~Hawking,
``Particle Creation by Black Holes,''
Commun. Math. Phys. \textbf{43} (1975), 199-220
[erratum: Commun. Math. Phys. \textbf{46} (1976), 206]
doi:10.1007/BF02345020



\bibitem{JLMS}
D.~L.~Jafferis, A.~Lewkowycz, J.~Maldacena and S.~J.~Suh,
``Relative entropy equals bulk relative entropy,''
JHEP \textbf{06} (2016), 004
doi:10.1007/JHEP06(2016)004
[arXiv:1512.06431 [hep-th]].

\bibitem{Mertens2019}
T.~G.~Mertens and G.~J.~Turiaci,
``Defects in Jackiw-Teitelboim Quantum Gravity,''
JHEP \textbf{08} (2019), 127
doi:10.1007/JHEP08(2019)127
[arXiv:1904.05228 [hep-th]].



\bibitem{Witten2020a}
E.~Witten,
``Deformations of JT Gravity and Phase Transitions,''
[arXiv:2006.03494 [hep-th]].

\bibitem{Witten2020b}
E.~Witten,
``Matrix Models and Deformations of JT Gravity,''
Proc. Roy. Soc. Lond. A \textbf{476} (2020) no.2244, 20200582
doi:10.1098/rspa.2020.0582
[arXiv:2006.13414 [hep-th]].

\bibitem{Mefford2020}
E.~Mefford and K.~Suzuki,
``Jackiw-Teitelboim quantum gravity with defects and the Aharonov-Bohm effect,''
JHEP \textbf{05} (2021), 026
doi:10.1007/JHEP05(2021)026
[arXiv:2011.04695 [hep-th]].

\bibitem{Fursaev2006}
D.~V.~Fursaev,
``Proof of the holographic formula for entanglement entropy,''
JHEP \textbf{09} (2006), 018
doi:10.1088/1126-6708/2006/09/018
[arXiv:hep-th/0606184 [hep-th]].

\bibitem{HarlowReview}
D.~Harlow,
``Jerusalem Lectures on Black Holes and Quantum Information,''
Rev. Mod. Phys. \textbf{88} (2016), 015002
doi:10.1103/RevModPhys.88.015002
[arXiv:1409.1231 [hep-th]].

\bibitem{Lashkari2018}
N.~Lashkari, H.~Liu and S.~Rajagopal,
``Modular Flow of Excited States,''
doi:10.1007/JHEP09(2021)166
[arXiv:1811.05052 [hep-th]].

\bibitem{Harlow2018}
D.~Harlow and D.~Jafferis,
``The Factorization Problem in Jackiw-Teitelboim Gravity,''
JHEP \textbf{02} (2020), 177
doi:10.1007/JHEP02(2020)177
[arXiv:1804.01081 [hep-th]].

\bibitem{Kitaev2017}
A.~Kitaev,
``Notes on $\widetilde{\mathrm{SL}}(2,\mathbb{R})$ representations,''
[arXiv:1711.08169 [hep-th]].

\bibitem{Trivedi2015}
R.~M.~Soni and S.~P.~Trivedi,
``Aspects of Entanglement Entropy for Gauge Theories,''
JHEP \textbf{01} (2016), 136
doi:10.1007/JHEP01(2016)136
[arXiv:1510.07455 [hep-th]].

\bibitem{Lin2018-2}
J.~Lin and \DJ{}.~Radi\v{c}evi\'c,
``Comments on defining entanglement entropy,''
Nucl. Phys. B \textbf{958}, 115118 (2020)
doi:10.1016/ j.nuclphysb.2020.115118
[arXiv:1808.05939 [hep-th]].


\bibitem{Skenderis2002}
K.~Skenderis,
``Lecture notes on holographic renormalization,''
Class. Quant. Grav. \textbf{19}, 5849-5876 (2002)
doi:10.1088/0264-9381/19/22/306
[arXiv:hep-th/0209067 [hep-th]].

\bibitem{Maldacena2016JT}
J.~Maldacena, D.~Stanford and Z.~Yang,
``Conformal symmetry and its breaking in two dimensional Nearly Anti-de-Sitter space,''
PTEP \textbf{2016} (2016) no.12, 12C104
doi:10.1093/ptep/ptw124
[arXiv:1606.01857 [hep-th]].


\bibitem{Botta2018}
M.~Botta-Cantcheff, P.~J.~Mart\'\i{}nez and G.~A.~Silva,
``The Gravity Dual of Real-Time CFT at Finite Temperature,''
JHEP \textbf{11} (2018), 129
doi:10.1007/JHEP11(2018)129
[arXiv:1808.10306 [hep-th]].

\bibitem{Botta2019}
M.~Botta-Cantcheff, P.~J.~Mart\'\i{}nez and G.~A.~Silva,
``Holographic excited states in AdS Black Holes,''
JHEP \textbf{04} (2019), 028
doi:10.1007/JHEP04(2019)028
[arXiv:1901.00505 [hep-th]].

\bibitem{Arias2020}
R.~Arias, M.~Botta-Cantcheff, P.~J.~Martinez and J.~F.~Zarate,
``Modular Hamiltonian for holographic excited states,''
Phys. Rev. D \textbf{102} (2020) no.2, 026021
doi:10.1103/PhysRevD.102.026021
[arXiv:2002.04637 [hep-th]].


\bibitem{Aitor2013}
A.~Lewkowycz and J.~Maldacena,
``Generalized gravitational entropy,''
JHEP \textbf{08} (2013), 090
doi:10.1007/JHEP08(2013)090
[arXiv:1304.4926 [hep-th]].

\bibitem{Rangamani2016}
M.~Rangamani and T.~Takayanagi,
``Holographic Entanglement Entropy,''
Lect. Notes Phys. \textbf{931} (2017), pp.1-246
doi:10.1007/978-3-319-52573-0
[arXiv:1609.01287 [hep-th]].

\bibitem{Nishioka2018}
T.~Nishioka,
``Entanglement entropy: holography and renormalization group,''
Rev. Mod. Phys. \textbf{90} (2018) no.3, 035007
doi:10.1103/RevModPhys.90.035007
[arXiv:1801.10352 [hep-th]].


\bibitem{Nakaguchi2016}
Y.~Nakaguchi and T.~Nishioka,
``A holographic proof of R\'enyi entropic inequalities,''
JHEP \textbf{12} (2016), 129
doi:10.1007/JHEP12(2016)129
[arXiv:1606.08443 [hep-th]].

\bibitem{Verlinde2019}
E.~Verlinde and K.~M.~Zurek,
``Spacetime Fluctuations in AdS/CFT,''
JHEP \textbf{04} (2020), 209
doi:10.1007/JHEP04(2020)209
[arXiv:1911.02018 [hep-th]].


\bibitem{Botta2014}
M.~Botta Cantcheff,
``Area Operators in Holographic Quantum Gravity,''
[arXiv:1404.3105 [hep-th]].

\bibitem{Jafferis2014}
D.~L.~Jafferis and S.~J.~Suh,
``The Gravity Duals of Modular Hamiltonians,''
JHEP \textbf{09} (2016), 068
doi:10.1007/JHEP09(2016)068
[arXiv:1412.8465 [hep-th]].


\bibitem{Yang2018}
Z.~Yang,
``The Quantum Gravity Dynamics of Near Extremal Black Holes,''
JHEP \textbf{05} (2019), 205
doi:10.1007/JHEP05(2019)205
[arXiv:1809.08647 [hep-th]].

\bibitem{Verlinde2020}
H.~Verlinde,
``ER = EPR revisited: On the Entropy of an Einstein-Rosen Bridge,''
[arXiv:2003.13117 [hep-th]].

\bibitem{Louis1993}
D.~Louis-Martinez, J.~Gegenberg and G.~Kunstatter,
``Exact Dirac quantization of all 2-D Dilaton gravity theories,''
Phys. Lett. B \textbf{321} (1994), 193-198
doi:10.1016/0370-2693(94)90463-4
[arXiv:gr-qc/9309018 [gr-qc]].

\end{thebibliography}
\end{document}